\documentclass{IEEEtran}
\usepackage{cite}
\usepackage{amsmath,amssymb,amsfonts}
\usepackage{algorithm}
\usepackage{algorithmic}
\usepackage{graphicx}
\usepackage[caption=false]{subfig}
\usepackage{textcomp}
\usepackage{xcolor}
\usepackage{bm}
\usepackage{empheq}
\usepackage{multirow}
\usepackage{hyperref}
\graphicspath{ {Images/} }
\def\BibTeX{{\rm B\kern-.05em{\sc i\kern-.025em b}\kern-.08em
    T\kern-.1667em\lower.7ex\hbox{E}\kern-.125emX}}

\usepackage[english]{babel}
\begin{document}
\renewcommand{\tablename}{TABLE}
\renewcommand{\refname}{REFERENCES}

\title{A Kalman Filter-Based Tracking Loop Design for Real-Time Aerospace GNSS Applications with Minimum Pull-Out Probability}

%==========================================================================
% TAES Article Author format (OPTIMIZED)
%==========================================================================
\author{Llorente, J. Fermín$^{\dagger *}$; Smidt, Javier A.; Roncagliolo, Pedro A.; López La Valle, Ramón
\vspace{0.8em}\\
\small{\textit{Electronic Navigation and Telecommunications Systems (SENyT) Research Group,}}\\
\small{\textit{Department of Electrical Engineering, School of Engineering, National University of La Plata (UNLP), La Plata, Argentina.}}\\
\small{$^\dagger$\textit{National Scientific and Technical Research Council (CONICET), La Plata 1900, Argentina.}}\\
\small{$^*$\texttt{\href{mailto:fermin.llorente@ing.unlp.edu.ar}{fermin.llorente@ing.unlp.edu.ar}}}
\thanks{This work has been submitted to the IEEE for possible publication. Copyright may be transferred without notice, after which this version may no longer be accessible.\\\textbf{Note:} This is the revised manuscript (Sep 2026). The original preprint (v1) was posted on TechRxiv (DOI: \href{https://doi.org/10.36227/techrxiv.176344172.21345819/v1}{10.36227/techrxiv.176344172.21345819/v1}).}
}

\maketitle

\begin{abstract}
    Kalman filter-based (KF-based) tracking loops are a powerful alternative to traditional phase-locked loops (PLLs) for Global Navigation Satellite Systems (GNSS) signal tracking. The primary advantage of the KF is its ability to incorporate high-fidelity models for receiver dynamics and clock errors, allowing the loop to adapt optimally to signal conditions. However, this theoretical optimality is often compromised in practice by the processing delays inherent in real-time systems with hardware correlators, which existing KF formulations typically neglect. This paper introduces a Modified Kalman filter (mKF) that overcomes this limitation specifically for hardware-based architectures. By reformulating the measurement update to be consistent with the processing delays, the proposed mKF maintains optimality in a practical implementation. We further present a systematic method for tuning both the process noise covariance matrix and the correlation time, based on an analytical expression for the pull-out probability (POP), which is validated through Monte Carlo simulation. The mKF is then validated with a GNSS signal simulator, both by post-processing baseband samples and on a real-time GPS receiver with hardware correlators. A direct equivalence between the mKF and a one-delay Digital PLL (DPLL) is established entirely in the digital domain. At equal noise bandwidth, the mKF matches the DPLL's phase error variance while achieving lower error in the higher-order states. Moreover, the mKF sustains lock at bandwidths inaccessible to the optimal one-delay DPLL under the same dynamic stress, positioning the proposed architecture as a robust and noise-efficient solution for high-dynamic aerospace GNSS applications.
\end{abstract}
\section{INTRODUCTION}\label{sec:intro}
    The core of a Global Navigation Satellite Systems (GNSS) receiver is the signal synchronization and tracking stage. Spread spectrum signals require synchronization to extract the navigation information. Code and carrier tracking loops perform this task. Since the receiver is expected to operate under challenging conditions, tracking loops need to be designed carefully to estimate the degraded or highly time-variant signal parameters. Specifically, for receivers in high-dynamic applications such as launch vehicles or sounding rockets, the dominant dynamic event is the acceleration step at liftoff (produced by engine ignition), modeled as a step in acceleration with effectively infinite jerk. The receiver acquires and locks the available signals during the static pre-launch phase, and the tracking loops must be designed to maintain lock throughout this acceleration transient.
    
    The most popular carrier tracking scheme is the Phase-Locked Loop (PLL) due to its low computational complexity. Although the PLL is the standard solution, its traditional design methodologies, often based on straightforward analog principles, suffer from a critical drawback: significant bandwidth limitations~\cite{book:kaplan2006}. This weakness has motivated alternative approaches, including maximum likelihood filtering \cite{article:hurd1987HighDynamicGPSreceiver} and native digital designs suitable for high-dynamics receivers~\cite{inproceedings:par2007ION}. The latter reference presents an optimum loop filter design method for a given input dynamic profile. This optimality is defined in terms of a weighted sum of the transient response energy and output noise power, enabling robust tracking under dynamic stress of up to $40$\,g.

    Since synchronization is the primary objective, clock errors play a crucial role in tracking loop performance~\cite{inproceedings:AVARandKF1984VanDie}. The so-called correlation stage requires precise timing to produce outputs that reflect the estimation errors. While clock dynamics can be characterized by Allan variance (AVAR)~\cite{article:allan1971FrequencyStability}, a typical measure used to specify the stability of oscillators over time, traditional PLLs primarily model thermal noise and receiver dynamics. Although adaptive bandwidth PLLs exist~\cite{article:Icortes2021AdaptivePLL}, they often incorporate clock dynamics in a heuristic manner~\cite{article:song2022TableBasedAdaptivePLL,inproceedings:khan2010projectedBWloop}, rather than through an optimization process.
 
    To fully integrate sophisticated clock models, like those derived from AVAR, directly into the tracking loop, a more powerful estimation framework is required. Within a state-space formulation, one of the most popular architectures is the Kalman filter (KF)~\cite{article:arePLLsDead,article:survey2014LopezSalcedo,article:wonPany2012}. This architecture exhibits an inherent adaptive nature and has been widely adopted for GNSS signal tracking due to its optimality under the minimum mean square error (MMSE) criterion. The main drawback of this architecture is the increase in computational burden. Also, considering implementation delays is crucial for real-time applications, as neglecting them can lead to instability or sub-optimal performance, especially in closed-loop schemes~\cite{article:patapoutian2002KalmanFilterLoopDelay}. While much of the existing literature implicitly assumes a zero-delay correlator model, often feasible in software-only implementations, this overlooks the critical timing constraints imposed by hardware correlators. This work aims to address the timing and delay constraints present in real-time implementations of KF-based tracking loops with hardware correlators. We propose a Modified KF (mKF) formulation that inherently accounts for these delays. 
    
    The theoretical optimality of the KF is contingent upon the precise tuning of the filter's parameters, a task that presents a significant practical challenge. While the problem of estimating the process noise covariance matrix has been studied and adaptive tuning methods for the KF have been proposed~\cite{inproceedings:vilavallsClosas2015IdentifNoiseStats,inproceedings:Icortes2021AdaptiveKF}, we present a systematic method for tuning the process noise covariance matrix to obtain optimal performance, defined in terms of a minimum tracking threshold and therefore minimum pull-out probability.

    The remainder of this paper is organized as follows. Section~\ref{sec:signalModel} describes the signal model for GNSS signal tracking. Section~\ref{sec:KFbasedTrk} presents the state-of-the-art in KF-based tracking, discusses the constraints of traditional formulations, and introduces the proposed mKF formulation that overcomes these timing limitations. In Section~\ref{sec:mKF}, the mKF is characterized, its performance is evaluated analytically, through simulation and with a real-time implementation in hardware, concluding with the proposed tuning method. An equivalence between the mKF and a one-delay Digital PLL (DPLL) is established in Section~\ref{sec:KFvsPLL}, including a performance comparison based on the tracking threshold. Finally, conclusions are drawn in Section~\ref{sec:conclusions}.
\section{SIGNAL MODEL}\label{sec:signalModel}
    A GNSS receiver operates by correlating the received signal with locally generated replicas in order to obtain the inputs of the digital signal processing chain. This is done in intervals of fixed duration and repeated continuously, resulting in an update rate equal to the inverse of the correlation time $T$. The correlation stage is usually implemented in hardware and the results are fed to the tracking loops, implemented in software, that will estimate signal parameters and configure local replica generators iteratively. The complex correlation in the \mbox{$k$-th} correlation interval (also called tracking epoch) for a given satellite with carrier-to-noise density ratio $C/N_0$ can be expressed as:
    \begin{equation}\label{eq:complexCorr1}
        r_k = d_k \sqrt{C/2} \,\mathrm{sinc}(\Delta f_kT)\,e^{j(\pi\Delta f_kT + \Delta\theta_k)} + \eta_k \ .
    \end{equation}
    The signal is modulated by binary data bits ($d_k=\pm 1$) using a Binary Phase Shift Keying (BPSK) scheme, which is a common feature in many GNSS signals, including legacy GPS signals and components of modernized ones. This work focuses exclusively on carrier tracking; therefore, the code delay tracking error is assumed to be zero and is omitted from the model (a condition met in practice when a carrier-aided code loop maintains sufficiently small residual code delay error~\cite{book:kaplan2006}). We define ${\mathrm{sinc}(x)=\sin(\pi x)/(\pi x)}$. The carrier frequency and phase error between the incoming signal, $s_k$, and the internal replica are $\Delta f_k = f_k - \hat{f}_k$ and $\Delta\theta_k = \theta_k - \hat{\theta}_k$, respectively. The term $\eta_k$ is a complex Gaussian random variable with zero mean and variance $\sigma^2 = N_0/(2T)$~\cite{book:spilker1996a}. The $C/N_0$ reported throughout this work refers to the input $C/N_0$ at the receiver, any sinc-induced attenuation from residual frequency tracking error reduces the effective post-correlation $C/N_0$.

    The tracking stage achieves synchronization lock when the difference between the received signal parameters and its locally generated replica converges to zero. This condition can be analyzed separately for phase and frequency, defining phase lock and frequency lock~\cite{book:chapter:springerHandbook2017signalProcessing}. In the operational scenarios considered in this work, the receiver acquires and locks all channels during a static pre-launch phase, and the algorithms presented are designed to maintain lock through the subsequent acceleration step at liftoff. Under these conditions, the $\mathrm{sinc}(\cdot)$ term in \eqref{eq:complexCorr1} can be approximated as unity, which is a suitable simplification for the subsequent analytical derivations, with the resulting $\mathrm{sinc}$ attenuation bounded to sub-decibel levels in the operating regime considered. All numerical simulations employ the full correlation model of \eqref{eq:complexCorr1}. By normalizing the signal term, the resulting approximated correlation can be modeled as:
    \begin{equation}\label{eq:complexCorr2}
        r_k = I_k +jQ_k \approx  d_k e^{j(\Delta\phi_k)} + n_{I_k} + jn_{Q_k} \ ,
    \end{equation}
    where the sequences $n_{I_k}$ and $n_{Q_k}$ are independent discrete time white Gaussian random processes with variance $\sigma^2 = 1/(2TC/N_0)$. Here, $\Delta\phi_k = \phi_k - \hat\phi_k$ represents the phase error at the middle of the correlation period, where the average phase in the interval is defined as $\phi_k = \pi f_kT + \theta_k$, with $\theta_k$ representing the signal phase at the beginning of the period. With the presented model, the loop can be treated as a purely digital single-input single-output (SISO) system~\cite{article:par2012}. The subscripts in~\eqref{eq:complexCorr1}--\eqref{eq:complexCorr2} denote the signal and replica parameters present during the correlation interval that produces~$r_k$; their precise temporal alignment with the filter's state vector depends on the tracking architecture and is made explicit in Section~\ref{ssec:mKF}.
\subsection{Phase Discriminator}
    Phase estimation error is calculated by feeding the correlation of~\eqref{eq:complexCorr2} to a discriminator. There are several options, but we will focus on the maximum likelihood estimator of the phase: the arctangent discriminator~\cite{book:spilker1996a},
    \begin{equation}\label{eq:phaseDetector}
        e_p[k] = \tan^{-1}\left( Q_k/I_k \right) = \left[ \Delta\phi_k + n_{\phi_k} \right]_\pi \ ,
    \end{equation}
    where the operator $\left[ \cdot \right]_\pi$ is responsible for keeping the value of $e_p[k]$ within the interval $(-\pi/2;\pi/2]$ of the $\tan^{-1}$ range. The zero-mean noise term $n_{\phi_k}$ has a non-trivial distribution due to the non-linearity of the discriminator, but under the assumption of adequate $C/N_0$ it can be approximated by a zero-mean Gaussian with variance \mbox{$\sigma^2_{\phi_k}\approx 1/(2TC/N_0)$}, neglecting the squaring-loss correction, which contributes less than $\sim0.4$\,dB of variance error in the operating regime considered~\cite{article:par2012,book:kaplan2006}. The choice of this discriminator is motivated by its insensitivity to signal amplitude and its efficient implementation via a lookup table, a method well-suited for the low-bit quantization typically applied to the components of the correlator output.

    The cyclic nature of this memoryless discriminator makes it unable to distinguish phase changes corresponding to integer cycles (or half-cycles when navigation data bits are present). This ambiguity can be resolved by adding a block with memory after the discriminator. One such solution is the Unambiguous Frequency Aided (UFA) block~\cite{inproceedings:par2007ION}, which operates as follows:
    \begin{equation}\label{eq:UFA}
        e_U[k] = e_p[k] - I_\pi\left( e_p[k] - e_U[k-1] \right) \ ,
    \end{equation}
    where $I_\pi\left(x\right) = x - \left[x\right]_\pi$. The algorithm is initialized with $e_U[0] = e_p[0]$. This addition can be interpreted as an extension of the linear range of the arctangent discriminator. As long as the frequency error remains within the range defined by the loop's update rate, cycle slips and pull-outs---and the long transient responses they cause---are avoided. 
    \subsection{State-Space Formulation}\label{sec:spaceStateFormulation}
    Considering a third order kinematic model, the three state vector is $\mathbf{x}_k = [ \phi_k , \omega_k , \dot{\omega}_k ]^T$. The average phase state, $\phi_k$, is the same as defined after~\eqref{eq:complexCorr2}, while the frequency and frequency rate states, $\omega_k$ and $\dot{\omega}_k$, are transformed to units of radians instead of cycles. The time reference of these three parameters is set at the center of the correlation interval because it represents an average estimate of the parameter in the current epoch.

    The time update of the state is straightforward considering discrete time integrations between the states. The state transition equation is:
    \begin{equation}\label{eq:dynamicEquation}
        \mathbf{x}_{k+1} = \mathbf{F}\mathbf{x}_k + \mathbf{w}_k \ ,
    \end{equation}
    where
    \begin{equation}\label{eq:F}
        \mathbf{F} = \begin{bmatrix}
            1 & T & T^2/2 \\
            0 & 1 & T \\
            0 & 0 & 1
        \end{bmatrix} \ .
    \end{equation}
    The process noise vector $\mathbf{w}_k = [w_{\phi},w_\omega,w_{\dot{\omega}}]_k^T$ is a zero-mean Gaussian vector with covariance matrix $\mathbf{Q}$. This matrix accounts for two distinct sources of error: uncertainty in the user's kinematics (those unmodeled by $\mathbf{F}$) and instability from the receiver's local oscillator~\cite{book:brown4ed2012,inproceedings:AVARandKF1984VanDie}.

    The kinematic component, $\mathbf{Q}_D$, arises from the assumption that changes in acceleration (jerk) can be modeled as a white noise process. If this continuous-time process has a power spectral density (PSD) $S_j$, the corresponding discrete-time covariance matrix for our third-order system is~\cite{book:brown4ed2012,article:1yadeMorton2017generalizedGnssSignalTRK}:
    \begin{equation}\label{eq:QmatrixDynamics}
        \mathbf{Q}_D = S_j\begin{bmatrix}
        T^5/20 & T^4/8 & T^3/6\\ 
        T^4/8 & T^3/3 & T^2/2\\ 
        T^3/6 & T^2/2 & T
        \end{bmatrix} \ .
    \end{equation}
    While $\mathbf{Q}_D$ effectively models unpredictable user motion, we wish to assess the impact of explicitly modeling clock errors. To this end, a second covariance term, $\mathbf{Q}_A$, is introduced.
    
    This second matrix, $\mathbf{Q}_A$, captures the stochastic noise characteristics of the oscillator, which are typically described using AVAR. It is constructed based on two underlying noise processes with PSD, $S_f$ and $S_g$, defined as follows~\cite{book:brown4ed2012},
    \begin{equation}\label{eq:QmatrixAVARparams}
        S_f = \dfrac{h_0}{2} \ , \ S_g = 2\pi^2h_{-2} \ . 
    \end{equation}
    For this work, we chose the parameters $h_0 = 1.8 \times 10^{-20}$ and $h_{-2} = 1.24 \times 10^{-21}$ to account for the phase noise specification of a standard temperature-compensated crystal oscillator (TCXO) of the GNSS receiver developed at SENyT~\cite{inproceedings:SR2017URUCONqseries}. These values are used in the analysis and characterization of the mKF, except in \mbox{Section~\ref{sec:mKF}-\ref{ssec:validationGSG8samples}} where it is explicitly stated. This leads to a 2$\times$2 covariance matrix for the clock states~\cite{inproceedings:AVARandKF1984VanDie},
    \begin{equation}\label{eq:QmatrixAVAR}
        \mathbf{Q}_A = \omega_{0}^2S_f\begin{bmatrix}
        T & 0\\ 
        0 & 0
        \end{bmatrix} + \omega_{0}^2S_g\begin{bmatrix}
        T^3/3 & T^2/2\\ 
        T^2/2 & T 
        \end{bmatrix} \ ,
    \end{equation}
    where $\omega_0$ is the nominal carrier angular frequency, set to the GPS L1 value $\omega_0 = 2\pi \times 1575.42 \times 10^6$~rad/s in this work. The final process noise matrix for the filter is formed by combining these two components:
    \begin{equation}\label{eq:QmatrixFull}
        \mathbf{Q} = \mathbf{Q}_D + 
        \begin{bmatrix}
            \mathbf{Q}_A & \bm{0}_{2\times 1}\\ 
            \bm{0}_{1\times 2} & 0
        \end{bmatrix} \ .
    \end{equation}
\section{KALMAN FILTER-BASED SIGNAL TRACKING}\label{sec:KFbasedTrk}
    KF-based architectures for signal tracking have been extensively studied by the GNSS community, leading to a variety of existing formulations. This interest is mainly due to the simplicity with which a state-space model can incorporate noise statistics and adaptivity (time variation). In our case, the state-space formulation allows designers to incorporate high-fidelity models for secondary error sources, such as clock dynamics. The KF's inherent adaptive nature allows selection of the bandwidth according to these three parameters: $C/N_0$, receiver dynamics, and clock dynamics. The correlation time $T$ can also be considered a design parameter; in this work it is fixed at $T = 5$\,ms, selected from the analytical pull-out probability evaluation over the full $(T, S_j)$ design space, as discussed in Section~\ref{sec:mKF}-\ref{ssec:popAnalysis}.

    The primary challenge in applying a KF to this problem lies in handling the nonlinear relationship between the state vector, $\mathbf{x}_k$, and the raw outputs from the correlation stage, modeled in~\eqref{eq:complexCorr1}. To address this efficiently, the literature widely adopts the direct-state KF~\cite{article:wonPany2012}. Instead of using the raw complex correlation values as measurements, this approach modifies the measurement model to consist simply of the true phase state plus additive noise. Under this formulation, the output of the phase discriminator, $e_p[k]$, acts directly as the innovation sequence $(\mathbf{y}_k - \hat{\mathbf{y}}_k^-)$ for a standard linear KF. Because the redefined measurement model is linear, it utilizes a constant measurement matrix. This completely avoids the computational burden and potential linearization inaccuracies associated with calculating a Jacobian matrix at every epoch, which would be required if applying an Extended Kalman Filter (EKF) directly to the raw correlators~\cite{book:brown4ed2012}.
    
    However, this direct-state formulation introduces a different challenge because the measurement noise is no longer strictly Gaussian. Under typical operating conditions with adequate $C/N_0$, this noise can be effectively approximated as a Gaussian process, an assumption that is standard in the KF-based tracking literature~\cite{article:wonPany2012,article:survey2014LopezSalcedo,article:arePLLsDead,article:1yadeMorton2017generalizedGnssSignalTRK}. The Gaussian approximation is analogous to those made in classical PLL design. This direct-state KF is considered a reference architecture due to its robust performance and implementation efficiency~\cite{article:survey2014LopezSalcedo}, and is therefore the approach adopted for the analysis in this work.
    
    The resulting state-space model is as follows,
    \begin{equation}\label{eq:stateSpaceModel}
        \left\{\begin{matrix}
        \mathbf{x}_{k+1} &= \mathbf{F}\mathbf{x}_k + \mathbf{w}_k    \\ 
        \mathbf{y}_k &= \mathbf{H}\mathbf{x}_k + \mathbf{v}_k  
        \end{matrix}\right. \ .
    \end{equation}
    The output matrix $\mathbf{H}$ sets the relationship between the states and the measurements; its form depends on the tracking architecture and is specified in Sections~\ref{sec:KFbasedTrk}-\ref{ssec:tradKF} and~\ref{sec:KFbasedTrk}-\ref{ssec:mKF}. The term $\mathbf{v}_k$ represents the measurement noise---assumed to be independent of the process noise $\mathbf{w}_k$---with a covariance matrix \mbox{$\mathbf{R} = 1/(2TC/N_0)$} (see Section~\ref{sec:signalModel}).

    The operation is the standard KF algorithm as described in~\cite{book:brown4ed2012}, with a-priori estimate defined as,
    \begin{equation}\label{eq:KFtimeUpdate}
        \hat{\mathbf{x}}_{k}^- = \mathbf{F}\hat{\mathbf{x}}_{k-1}^+ \ ,
    \end{equation}
    and a-posteriori estimate as,
    \begin{equation}\label{eq:KFupdate}
        \hat{\mathbf{x}}_{k}^+ = \hat{\mathbf{x}}_{k}^- + \mathbb{K}_k\left( \mathbf{y}_k - \mathbf{H}\hat{\mathbf{x}}_{k}^- \right) \ ,
    \end{equation}
    where $\mathbb{K}_k$ is the Kalman gain matrix. To ensure the measurement residual, $(\mathbf{y}_k - \mathbf{H}\hat{\mathbf{x}}_{k}^-)$, is equivalent to the phase discriminator output, it is crucial to set the local replica's reference using the a-priori estimate $\hat{\mathbf{x}}_k^-$~\cite{article:arePLLsDead}.
\subsection{Traditional KF Formulations}\label{ssec:tradKF}  
    The challenge of processing delays is well documented in the context of digital tracking loops. Traditional analog-based designs often ignore this issue. However, in high-dynamic contexts where wide bandwidths are required, discretizing an analog model using standard approximations yields poor results. Specifically, as the product of the noise equivalent bandwidth and the update interval ($B_N T$) approaches unity, the actual digital bandwidth diverges significantly from the analog design intent, eventually leading to instability when closing the loop~\cite{book:kaplan2006}. To overcome this issue, digital PLL design methods have been proposed that explicitly account for both the correlation delay and the subsequent processing time~\cite{article:par2012}. However, the literature on KF-based tracking largely overlooks the timing constraints inherent in practical, hardware-based implementations. This is often because many formulations implicitly assume a zero-delay correlator model, which, while feasible in software-only closed loops, is not valid for receivers with hardware correlators. A proper model for these delays is critical, as ignoring them exacerbates the aforementioned stability issues~\cite{article:bergmans1995EffectLoopDelayDPLL}.

    Direct-state KF-based tracking loops typically use a measurement matrix, $\mathbf{H} = \left[ 1 \ 0 \ 0 \right]$, which selects the phase state. This results in an innovation term, $(\mathbf{y}_k - \mathbf{H}\hat{\mathbf{x}}_k^-)$, that is equivalent to the discriminator's output. While this is a common approach~\cite{article:survey2014LopezSalcedo,article:wonPany2012}, other formulations exist, such as those that propagate the states to form a mean phase error~\cite{article:1yadeMorton2017generalizedGnssSignalTRK}. Both formulations are largely equivalent, but differ in the temporal alignment of the state estimates.

\begin{figure}
        \subfloat[Traditional KF-based tracking loop, ignoring the processing delay.]{
            \includegraphics[width=0.49\textwidth]{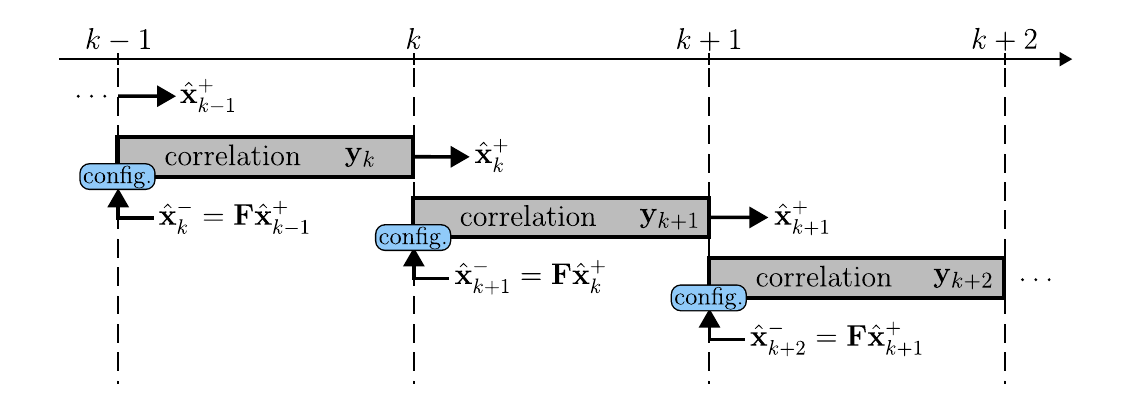}
            \label{fig:timesTradKF}
        }

        \subfloat[Modified KF proposed implementation, contemplating the processing delay.]{
            \includegraphics[width=0.49\textwidth]{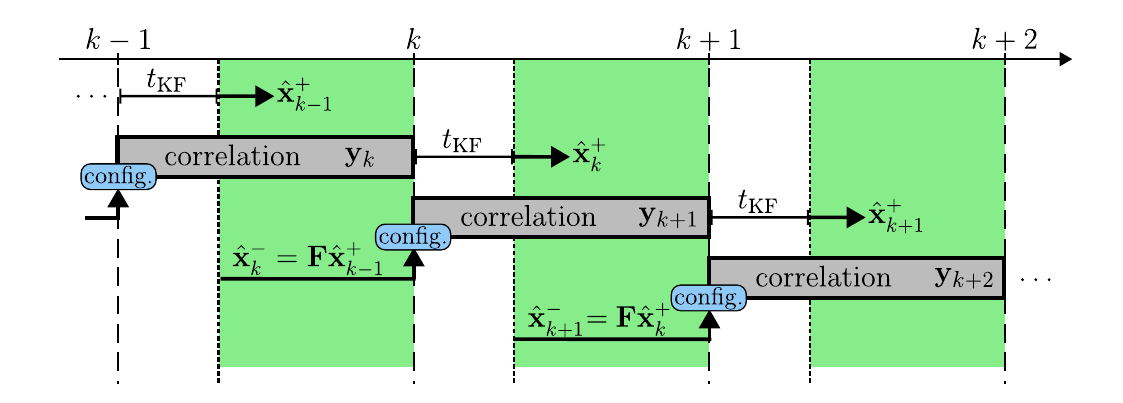}
            \label{fig:timesmKF}
        }
        \caption{Time diagram of the traditional and proposed KF-based tracking loops.}
        \label{fig:timesDiagramBoth}
    \end{figure}

\begin{figure}
        \subfloat[Traditional KF-based tracking loop.]{
            \includegraphics[width=0.49\textwidth]{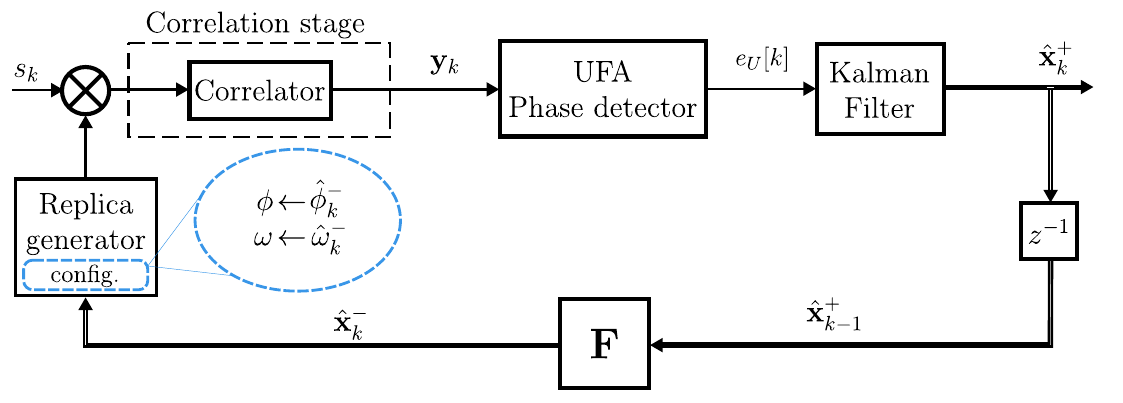}
            \label{fig:tradKFblockDiagram}
        }

        \subfloat[Proposed Modified KF-based tracking loop.]{
            \includegraphics[width=0.49\textwidth]{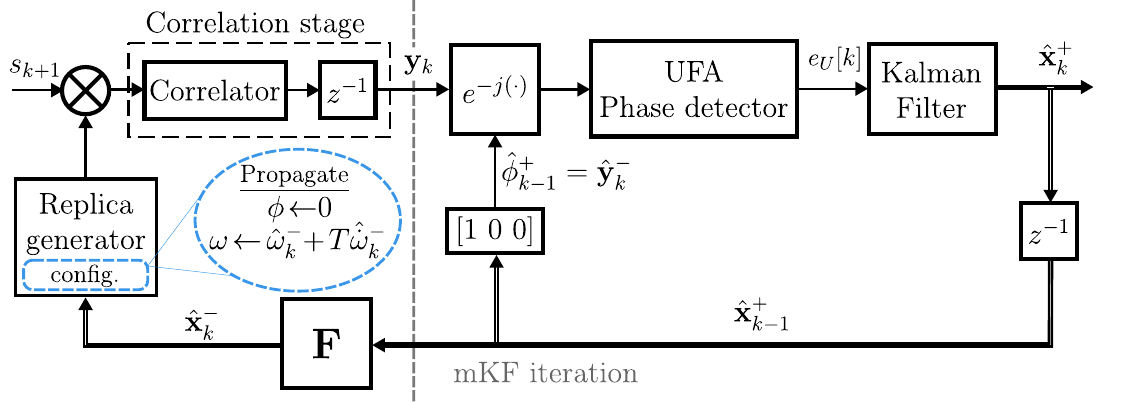}
            \label{fig:mKFblockDiagram}
        }
        \caption{Block diagrams of the traditional and proposed KF-based tracking loops.}
        \label{fig:kfBlockDiagramBoth}
    \end{figure}

    In GNSS receivers with hardware correlators, the phase and frequency used to generate the local replica must be set before the correlation for epoch $k$ begins. We define the aforementioned correlation as the product that is available at epoch $k$, from which $e_U[k]$ will be computed. As dictated by the KF recursion, this requires availability of the a-priori estimate, $\hat{\mathbf{x}}_{k}^-$, which is computed from the previous a-posteriori estimate, $\hat{\mathbf{x}}_{k-1}^+$. However, as illustrated in Fig.~\ref{fig:timesTradKF}, $\hat{\mathbf{x}}_{k-1}^+$ becomes available precisely at the moment the computation of the correlation for epoch $k$, $\mathbf{y}_k$, must begin. Therefore, the ideal procedure, illustrated in Fig.~\ref{fig:tradKFblockDiagram}, is only valid under an assumption of zero processing delay, where the KF update is computed instantaneously.
    
    In any practical, real-time implementation, this update requires a non-zero processing time, $t_{\text{KF}}$. This means the required estimate, $\hat{\mathbf{x}}_{k-1}^+$, is only actually available after the $k$-th correlation has already begun, as the correlators must run continuously to avoid losing portions of the signal. This timing conflict makes it impossible to configure the correlators with the true a-priori estimate as the ideal literature model suggests.
       
    Even though $t_{\text{KF}}$ is many orders of magnitude smaller than the loop update time $T$, it is not negligible. The computations required for the phase discriminator and state update can take dozens of clock cycles in standard processors. For a processor operating at tens of megahertz, this produces a time delay comparable to a C/A code chip period, which can translate to hundreds of meters of code tracking error~\cite{book:spilker1996a}. This code error could be mitigated by estimating $t_{\text{KF}}$ and compensating the code loop. However, the carrier tracking loop still loses a portion of the signal, leading to untracked carrier cycles. This can cause unavoidable cycle slips---which are fatal for high-precision positioning---in high-dynamics scenarios.

    Regardless of whether the excess delay is attributed to the correlation process or the subsequent computation, the critical point is that these timing restrictions make the ideal KF update from~\eqref{eq:KFupdate} unachievable. The use of a KF under loop delays results in a sub-optimal solution~\cite{article:patapoutian2002KalmanFilterLoopDelay}. This presents a difficult trade-off: the implementation complexity increases because of the KF architecture, while its primary advantage---optimality---is compromised. This questions the overall benefit of traditional KF solutions for real-time receivers with hardware correlators and motivates the search for alternative designs that can overcome this inherent processing delay.
\subsection{Proposed Modified KF Formulation}\label{ssec:mKF}
    As established, the processing delays inherent in the loop cannot be avoided and must be accounted for in the filter design. We propose a formulation that redefines the measurement update to compare the current measurement at epoch $k$ with the state estimate from the previous epoch, $k-1$. This is achieved by defining an output matrix that effectively propagates the state backwards by one time period, $T$, using the inverse of the state transition matrix from~\eqref{eq:F}. This results in:
    \begin{equation}\label{eq:H1r}
        \mathbf{H} = \bm{e}_1\mathbf{F}^{-1} = \left[1 \ \ -T \ \ T^2/2\right] \ ,
    \end{equation}
    where $\bm{e}_1 = \left[1 \ 0 \ 0 \right]$. This choice redefines the a-priori measurement estimate as,
    \begin{align}\label{eq:yest}
        \hat{\mathbf{y}}_k =&\mathbf{H}\hat{\mathbf{x}}_k^- = \bm{e}_1\hat{\mathbf{x}}_{k-1}^+ = \hat{\phi}_{k-1}^+ \ ,
    \end{align}
    where the result follows from the identity $\mathbf{F}^{-1}\mathbf{F} = \mathbf{I}$. The innovation term in~\eqref{eq:KFupdate} now compares the current measurement $\mathbf{y}_k$ with the previous a-posteriori phase estimate $\hat{\phi}_{k-1}^+$, which is available while the \mbox{$k$-th} correlation is being computed (see Fig.~\ref{fig:timesmKF}).

    Under the mKF timing, the correlator at epoch~$k$ uses the a-posteriori phase reference $\hat{\phi}_{k-1}^{+}$ from the previous epoch. Specializing the general signal model~\eqref{eq:complexCorr1}--\eqref{eq:phaseDetector} to this timing yields
    \begin{align}
        r_k &= d_{k-1}\sqrt{C/2}\;\mathrm{sinc}(\Delta f_{k-1}T)\;
               e^{j(\pi\Delta f_{k-1}T + \Delta\theta_{k-1})} + \eta_k,
        \label{eq:corr1mKF} \\
        r_k &\approx d_{k-1}\;e^{j\Delta\phi_{k-1}} + n_{I_k} + jn_{Q_k},
        \label{eq:corr2mKF}
    \end{align}
    where $\Delta f_{k-1} = f_{k-1} - \hat{f}_{k-1}$ and $\Delta\phi_{k-1} = \phi_{k-1} - \hat{\phi}_{k-1}^{+}$. The arctangent discriminator then gives the measurement fed to the filter (after UFA unwrapping):
    \begin{equation}\label{eq:eUmKF}
        e_U[k] = \bigl[\Delta\phi_{k-1} + n_{\phi_k}\bigr]_\pi = \bigl[\phi_{k-1} - \hat{\phi}_{k-1}^{+} + n_{\phi_k}\bigr]_\pi,
    \end{equation}
    which carries information about the previous-epoch phase $\phi_{k-1}$, not $\phi_k$. 

    The practical implementation of this method requires addressing how to configure the correlators for epoch~$k$. From~\eqref{eq:corr1mKF}, the complex correlation can be decomposed as
    \begin{align}\label{eq:corr1Rfix}
        r_k &= d_{k-1}\,\mathrm{sinc}(\Delta f_{k-1})\,
               e^{j\Delta\phi_{k-1}} + \eta_k \nonumber\\
            &= \bigl(d_{k-1}\,\mathrm{sinc}(\Delta f_{k-1})\,
               e^{j\phi_{k-1}} + \eta_k'\bigr)\,
               e^{-j\hat{\phi}_{k-1}^{+}} \ ,
    \end{align}
    where $\eta_k' = \eta_k\,e^{j\hat{\phi}_{k-1}^{+}}$ has the same distribution as $\eta_k$~\cite{inproceedings:par2016ION}. This shows that the phase reference $\hat{\phi}_{k-1}^{+}$ can be set to zero during correlation and its rotation applied digitally to $r_k$ post-correlation, provided the frequency reference is accurate.
    
    The frequency reference must be set in advance. However, as previously discussed, the a-priori frequency estimate for the upcoming epoch is not yet available when configuring the correlation stage. Instead, the available a-priori estimate from the current epoch can be propagated to approximate the frequency for the upcoming accumulation period. To clarify this timing relationship: at the moment of configuring the correlators for measurement $\mathbf{y}_{k+1}$, the most up-to-date state estimate available is $\hat{\mathbf{x}}_k^-$. This estimate is temporally misaligned with the signal frequency present during the correlation of $\mathbf{y}_{k+1}$, because $\hat{\omega}_k^-$ corresponds to the midpoint of the preceding correlation interval, $[(k-1)T,kT]$. To compensate for this, a simple linear prediction serves as a sufficiently accurate substitute for the true a-priori value ($\omega \leftarrow \hat{\omega}_{k}^- + T\hat{\dot{\omega}}_{k}^-$). This frequency estimator can be shown to be unbiased when tracking frequency rate steps. Functionally, this prediction mirrors the time-update step inherent in a traditional KF~\cite{article:arePLLsDead} (see Fig.~\ref{fig:tradKFblockDiagram}), applied one additional time to the frequency state exclusively for the purpose of correlator configuration. As the mKF adds one extra delay to the processing compared to the traditional KF, the frequency estimate needs to be propagated one extra epoch in the mKF to be able to configure the correlators.
    
    In contrast, the phase state within the mKF recursion exhibits only a single-epoch delay, as depicted on the right side of the block diagram in Fig.~\ref{fig:mKFblockDiagram}. Conceptually, this architecture separates the timing of how each state estimate is applied to the hardware: the phase estimate compensates the correlation output with a single-epoch delay, while the frequency estimate steers the local replica two epochs ahead. The KF itself, however, still updates all three states jointly from the same phase innovation at every epoch. This strategy, illustrated in Fig.~\ref{fig:timesmKF}, Fig.~\ref{fig:mKFblockDiagram}, and summarized in Algorithm~\ref{alg:mKFrecursion}, successfully resolves the timing conflict.

    \begin{algorithm}[]
        \caption{Per-epoch mKF recursion at epoch $k$.}
        \label{alg:mKFrecursion}
        \begin{algorithmic}[1]
            \REQUIRE $\hat{\mathbf{x}}_{k}^-$,\; $\mathbf{y}_k$ from correlators
            \STATE Rotate $\mathbf{y}_k$ by $-\hat{\phi}_{k-1}^+$ to obtain $e_U[k]$
            \STATE $\hat{\mathbf{x}}_k^+ \gets \hat{\mathbf{x}}_k^- + \mathbb{K}\, e_U[k]$
            \STATE $\hat{\mathbf{x}}_{k+1}^- \gets \mathbf{F}\,\hat{\mathbf{x}}_k^+$
            \STATE Set correlator frequency for $\mathbf{y}_{k+2}$:\\ \hspace{\algorithmicindent} $\omega \gets 2\pi(\hat{f}_{k+1}^- + T\,\hat{\dot{f}}_{k+1}^-)$
            \STATE \textbf{Meanwhile:} $\mathbf{y}_{k+1}$ is being accumulated
            \ENSURE $\hat{\mathbf{x}}_k^+$,\; $\hat{\mathbf{x}}_{k+1}^-$
        \end{algorithmic}
    \end{algorithm}

    Our proposed algorithm, the Modified KF (mKF), thus compares the incoming signal's phase with the \mbox{one-epoch-old} a-posteriori estimate, directly accounting for the unavoidable processing delay introduced by the hardware correlators. This results in an optimal KF for carrier tracking that is robust to implementation delays when closing the loop.

    \subsubsection*{Riccati recursion with cross-correlated noise}

    The choice $\mathbf{H} = \bm{e}_1\mathbf{F}^{-1}$ is an algebraic reformulation of the delayed observation $\mathbf{y}_k = \bm{e}_1\mathbf{x}_{k-1} + v_k$ into the standard form $\mathbf{y}_k = \mathbf{H}\mathbf{x}_k + \tilde{v}_k$. Under this reformulation, the effective measurement noise $\tilde{v}_k = v_k - \bm{e}_1\mathbf{F}^{-1}\mathbf{w}_{k-1}$ includes a process-noise contribution, creating a correlation between $\mathbf{w}_{k-1}$ and $\tilde{v}_k$. This cross-correlation does not affect the computed innovation, since $\mathbf{H}\hat{\mathbf{x}}_k^{-} = \bm{e}_1\mathbf{F}^{-1}\mathbf{F}\hat{\mathbf{x}}_{k-1}^{+} = \hat{\phi}_{k-1}^{+}$ carries no process noise realisation, but it modifies the Riccati recursion used to compute the Kalman gain. Following the standard treatment for correlated process and measurement noise~\cite{book:barShalom2004}, the cross-covariance is
    \begin{align}
        \mathbf{N}   &= -\mathbf{Q}\mathbf{F}^{-\top}\bm{e}_1^\top,
        \label{eq:Ncorr}
    \end{align}
    and the corrected gain and covariance recursion is
    \begin{align}
        \mathbf{P}_k^{-} &= \mathbf{F}\mathbf{P}_{k-1}^{+}\mathbf{F}^\top
                            + \mathbf{Q},
        \label{eq:Pminus}\\
        \mathbf{K}_k     &= \frac{\mathbf{P}_k^{-}\mathbf{H}^\top
                            + \mathbf{N}}
                            {\bm{e}_1\mathbf{P}_{k-1}^{+}\bm{e}_1^\top + R},
        \label{eq:Kcorr}\\
        \mathbf{P}_k^{+} &= \mathbf{P}_k^{-}
                            - \mathbf{K}_k
                            \bigl(\mathbf{F}\mathbf{P}_{k-1}^{+}
                            \bm{e}_1^\top\bigr)^\top.
        \label{eq:Pplus}
    \end{align}
    The identity $\mathbf{P}_k^{-}\mathbf{H}^\top + \mathbf{N} = \mathbf{F}\mathbf{P}_{k-1}^{+}\bm{e}_1^\top$ can be verified algebraically, confirming that the cross-correlation $\mathbf{N}$ exactly cancels the $\mathbf{Q}$-dependent term introduced by $\mathbf{H}^\top = \mathbf{F}^{-\top}\bm{e}_1^\top$ in the gain numerator. At steady state, the corrected gains are numerically equivalent to those obtained without the cross-correlation terms across all operative $(S_j, C/N_0)$ pairs.

    It is worth noting that the corrected recursion~\eqref{eq:Pminus}--\eqref{eq:Pplus} is the exact Riccati equation for the observation model $\mathbf{y}_k = \bm{e}_1\mathbf{x}_{k-1} + v_k$. At the DARE fixed point, it produces gains numerically equivalent to those of the optimal delayed-measurement Kalman filter, also known as one-epoch smoothing KF~\cite{book:barShalom2004}~(Ch.~5), confirming that the $\mathbf{H} = \bm{e}_1\mathbf{F}^{-1}$ formulation achieves optimal delayed-measurement performance while preserving the single-step recursion structure used in Section~\ref{sec:mKF}.
\section{MODIFIED KALMAN FILTER CHARACTERIZATION AND TUNING}\label{sec:mKF}
    In practice, the system model is never perfectly known and contains intrinsic approximations. Consequently, the noise covariance matrices ($\mathbf{Q}$ and $\mathbf{R}$) are often treated as tuning parameters to account for this model mismatch while simultaneously defining the filter's bandwidth. The selection of these tuning parameters is arguably the most critical step in the design of the loop. Common methods presented in the literature often rely on rule-of-thumb techniques or empirical scaling of the matrices.

    While the measurement noise covariance $\mathbf{R}$ can be estimated from the receiver's $C/N_0$ estimate~\cite{article:falletti2011LowComplexityCn0estimators}, and the structure of $\mathbf{Q}$ is defined, the scale factor $S_j$ needs to be set and is the main tuning parameter of the filter. This section develops an analytical framework for optimally selecting this parameter.
\subsection{Steady-State Transfer Function}
    Although the KF recursively computes its gain matrix, under certain conditions, this gain will converge to a constant, steady-state value. It can be shown that a unique, steady-state Kalman gain matrix exists, provided that the system is both observable (the $(\mathbf{F}, \mathbf{H})$ pair) and controllable (the $(\mathbf{F}, \mathbf{C})$ pair, where $\mathbf{C}$ is the Cholesky factor of $\mathbf{Q}$)~\cite{book:barShalom2004}. As both conditions are satisfied for this problem, we can analyze the filter's steady-state behavior. Based on~\eqref{eq:KFupdate}, the state update equations in this steady-state regime are: 
    \begin{empheq}[left=\empheqlbrace]{align}\label{eq:KFupdateSteadyState}
    \hat{\phi}_k^+ &= \hat{\phi}_{k-1}^+ + T\hat{\omega}_{k-1}^+ + \dfrac{T^2}{2}\hat{\dot{\omega}}_{k-1}^+ &+ k_1\epsilon_k & \nonumber\\
    \hat{\omega}_k^+ &= \hat{\omega}_{k-1}^+ + T\hat{\dot{\omega}}_{k-1}^+ &+ k_2\epsilon_k & \\
    \hat{\dot{\omega}}_k^+ &= \hat{\dot{\omega}}_{k-1}^+ &+ k_3\epsilon_k & \nonumber \ ,
    \end{empheq}
    where the innovation $\epsilon_k$ is given by:
    \begin{equation}\label{eq:epsilonK}
        \epsilon_k = \mathbf{y}_k - \hat{\mathbf{y}}_k^- = \mathbf{y}_k - \mathbf{H}\hat{\mathbf{x}}_k^- = \mathbf{y}_k - \hat{\phi}_{k-1}^+ \ .
    \end{equation}
    To derive the filter's transfer function, we take the \mbox{z-transform} of these equations. After algebraic manipulation, the a-posteriori phase estimate in the z-domain is found to be,
    \begin{equation}\label{eq:phiZ}
        \hat{\Phi}(z) = E(z)N(z) \ , 
    \end{equation}
    where $N(z)$ is the open-loop transfer function,
    \begin{align}\label{eq:Nz}
       N(z) =  \frac{k_1}{(1-z^{-1})} + \\
        &\frac{(Tk_2 + T^2/2k_3z^{-1})}{(1-z^{-1})^2} + \frac{T^2k_3z^{-2}}{(1-z^{-1})^3} \ . \nonumber
    \end{align}
    The innovation $E(z)$ can be found by transforming~\eqref{eq:epsilonK}. The noiseless measurement at epoch~$k$ is the previous-epoch phase $\phi_{k-1}$, represented in the z-domain by $\Phi(z)z^{-1}$. This yields:
    \begin{equation}\label{eq:epsilonZ}
        E(z) = \Phi(z)z^{-1} - \hat{\Phi}(z)z^{-1} + V(z) \ .
    \end{equation}
    Combining these results gives the closed-loop transfer function, $T(z)$,
    \begin{equation}\label{eq:phizAndTz}
        \hat{\Phi}(z) = \frac{\left[ \Phi(z) + V(z) \right]z^{-1}N(z)}{\left(1+z^{-1}N(z)\right)} = T(z) \Phi(z) + T(z) V(z) . 
    \end{equation}
    This methodology is adapted from \cite{article:shu2013ThirdOrderKalmanFilter} for our mKF formulation. The proposed filter's transfer function includes a delay that cannot be addressed by the traditional formulation.
    
    To connect the filter's performance to its tuning parameters, we derive the PSD of the measurement $\mathbf{y}_k$. The transfer function of the system without considering the measurement noise is $\mathbf{G}(z) = \mathbf{H}\left( z\mathbf{I}-\mathbf{F} \right)^{-1}$. As the physical measurement noise $v_k$ and the process noise $\mathbf{w}_k$ are independent, the cross-correlation terms are zero, and the output PSD is
    \begin{align}\label{eq:SyyPSD}
        &S_{yy}(z) = \mathbf{G}(z^{-1})\mathbf{Q}\mathbf{G}(z)^T + R \nonumber \\
        =& \frac{R}{(z-1)^6}\left[ z^6 - \alpha_1z^5 + \alpha_2z^4 - \alpha_3z^3 + \alpha_2z^2 - \alpha_1z + 1\right] \nonumber \\
        =& R\frac{(z-z_0)(z-z_1)(z-z_2)(z-z_0^{-1})(z-z_1^{-1})(z-z_2^{-1})}{(z-1)^6} \nonumber \\
        =& R\frac{(z^2-r_0z + 1)(z^2-r_1z + 1)(z^2-r_2z + 1)}{(z-1)^6}
    \end{align}
    where the coefficients $\alpha_i$ are functions of the KF tuning parameters,
    \begin{equation}\label{eq:alphas}
    \begin{split}
        &\alpha_1 = 6 + \frac{1}{R}\left(\frac{31}{120}S_jT^5 +  \frac{5}{6}S_gT^3 + S_fT \right)\\
        &\alpha_2 = 15 + \frac{1}{R}\left(\frac{47}{60}S_jT^5 + \frac{13}{3}S_gT^3 +4S_fT \right)\\
        &\alpha_3 = 20 + \frac{1}{R}\left(\frac{41}{20}S_jT^5 +            7S_gT^3 +6S_fT \right)\\
    \end{split}
    \end{equation}
    and $r_i = z_i + z_i^{-1}$ with ($i=0,1,2$) and $z_i$ defined in \eqref{eq:SyyPSD}. From~\eqref{eq:SyyPSD} the following relationship between $\alpha_i$ and $r_i$ is obtained:
    \begin{equation}\label{eq:alphasANDris}
        \left\{\begin{array}{rl}
            r_0 + r_1 + r_2 & =\; \alpha_1 \\
            r_0r_1 + r_0r_2 + r_1r_2 & =\; \alpha_2-3 \\
            r_0r_1r_2 & =\; \alpha_3 - 2\alpha_1
        \end{array}\right. .
    \end{equation}
    The next step is to perform a spectral factorization of $S_{yy}(z)$ to find its minimum-phase factor, which is a key step in relating the KF to the optimal Wiener filter (WF). This factorization takes the form $S_{yy}(z) = B(z)B(z^{-1})$, where $B(z)$ is:
    \begin{equation}\label{eq:BzSpectralFactSyy}
        B(z) = \sqrt{R} \frac{(z-z_0)(z-z_1)(z-z_2)}{(z-1)^3} \ .
    \end{equation}
    This minimum-phase condition restricts the roots $z_i$ to lie inside the unit circle, which is essential for a causal and stable filter. By stating the equivalence between a \mbox{steady-state} KF and a WF, the resulting closed-loop transfer function can be expressed as~\cite{article:1yadeMorton2017generalizedGnssSignalTRK},
    \begin{equation}\label{eq:transferFuncWF}
        \begin{aligned}
            T^{\text{WF}}(z) & = 1 - \frac{\sqrt{R}}{B(z)} \\
            & = \frac{z^2(3-z_s) + z(z_d-3) + (1-z_p)}{z^3-z_sz^2+z_dz-z_p}
        \end{aligned}
        \ ,
    \end{equation}
    where $z_s = z_0 + z_1 + z_2$, $z_d = z_0z_1 + z_0z_2 + z_1z_2$ and $z_p = z_0z_1z_2$. The open-loop transfer function results:
    \begin{align}\label{eq:Nzfunctionofzk}
       N(z) = \frac{(3-z_s)+(z_d-3)z^{-1} + (1-z_p)z^{-2}}{(1-z^{-1})^3} \ ,
    \end{align}
    allowing the comparison with the transfer function derived from the state-space equations of~\eqref{eq:Nz}. Through this we find a direct relationship between the steady-state Kalman gains ($k_i$) and the system coefficients ($z_s$, $z_d$ and $z_p$).
    \begin{equation}\label{eq:ZsAndKs}
        \left\{\begin{array}{rl}
            z_s & =\; 3-k_1 \\
            z_d & =\; k_3T^2/2 + Tk_2 - 2k_1 + 3 \\
            z_p & =\; -k_3T^2/2 + k_2T - k_1 + 1
        \end{array}\right. .
    \end{equation}
    In order to have an expression for the poles of the KF's closed-loop transfer function ($z_0$, $z_1$ and $z_2$), the roots of the denominator of~\eqref{eq:transferFuncWF} must be computed. Solving the system from~\eqref{eq:ZsAndKs} with the definitions of $z_s$, $z_d$ and $z_p$ in terms of $z_0$, $z_1$ and $z_2$, we obtain the KF gains as a function of the poles. This procedure is based on \cite{article:1yadeMorton2017generalizedGnssSignalTRK}, but the present work focuses on the mKF formulation. Whether solved algebraically or numerically, this procedure links the physical tuning parameters of the KF to its transfer function poles and zeros, which define the filter's behavior. 

\subsection{Noise Equivalent Bandwidth}
    \begin{figure}[!t]
        \centering
        \includegraphics[width=0.49\textwidth]{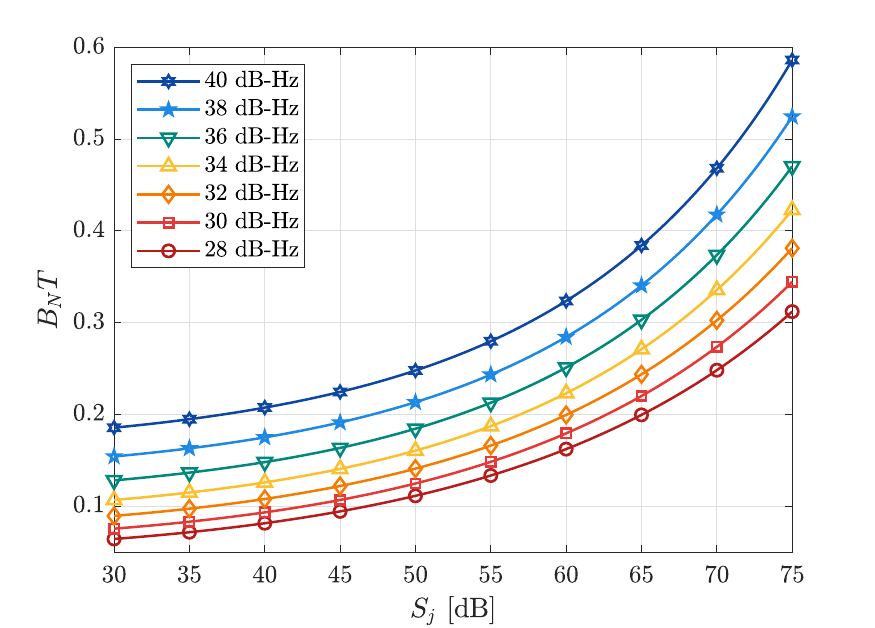}
        \caption{Normalized $B_N$ of the mKF as a function of $S_j$, for different $C/N_0$.}
        \label{fig:BnTmKF}
    \end{figure}
    \begin{figure}[!t]
        \centering
        \includegraphics[width=0.49\textwidth]{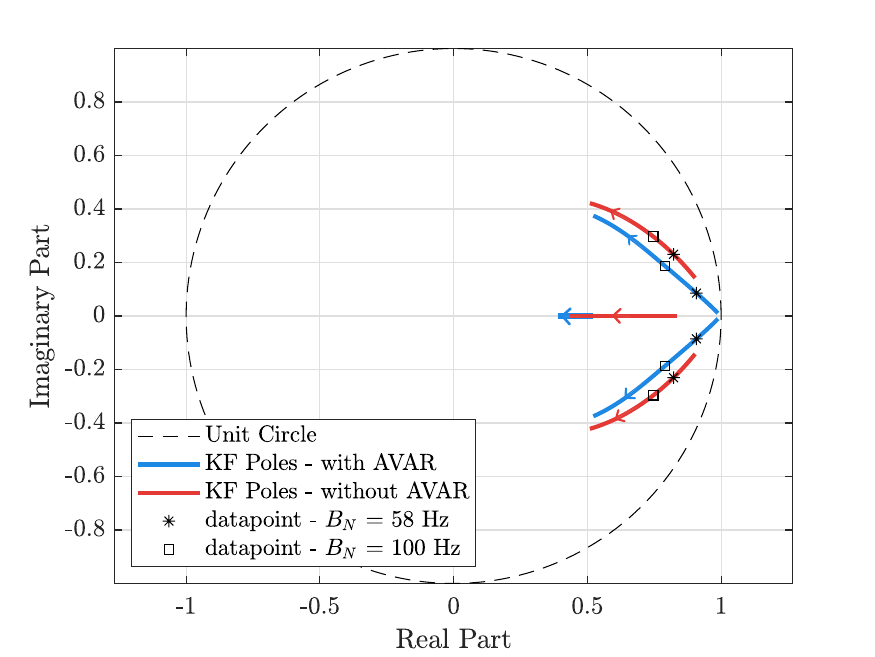}
        \caption{Root locus of the mKF as a function of $B_N$, comparing the models with and without the clock dynamics ($\mathbf{Q}_A$) component. The datapoints indicate the dominant poles.}
        \label{fig:rootLocusmKF}
    \end{figure}
    The noise equivalent bandwidth ($B_N$) is a common metric for evaluating tracking-loop performance. It can be computed from the closed-loop transfer function (Eq.~\eqref{eq:phizAndTz}) using the residue theorem,
    \begin{equation}\label{eq:kfBnT}
        B_NT = \frac{1}{2}\sum\limits_{i=0}^{2}\left\{ \text{Res}\left(T(z)T(z^{-1})z^{-1},z_i\right)\right\} \ .
    \end{equation}
    While the resulting expression is intricate, computing it numerically allows for the evaluation of the filter bandwidth for different tuning settings, as seen in Fig.~\ref{fig:BnTmKF}. Previous works often obtain the KF's $B_N$ by computing its equivalent DPLL bandwidth through the relationship with analog PLLs~\cite{article:won2013TuningMethodAdaptivePLLusingCRB,inproceedings:Icortes2021AdaptiveKF}, which can lead to discrepancies between the design bandwidth and the resulting digital bandwidth. As this work focuses on purely digital designs, the presented approach provides a means to calculate the true bandwidth of the KF-based tracking loop and enables designs that go beyond the traditional stability restriction $B_NT < 0.1$~\cite{book:kaplan2006}.

    The root locus of the mKF's closed-loop transfer function is also analyzed, with $B_N$ as the varying parameter. Fig.~\ref{fig:rootLocusmKF} compares the locus with and without the inclusion of clock dynamics in the filter model. While the overall structure of the pole locations is qualitatively similar in both scenarios, including the clock dynamics model consistently shifts the dominant poles closer to the unit circle (i.e., they exhibit greater magnitude) for any given $B_N$, as indicated by the datapoints in the figure. Complementing the results in \cite{inproceedings:jfl2024}---although obtained within a different framework---this demonstrates that modeling clock dynamics via AVAR alters the filter's behavior. Specifically, for the same level of platform dynamics ($S_j$), including clock noise results in a wider bandwidth ($B_N$). Conversely, tuning the filter to achieve the same $B_N$ when clock dynamics are included results in poles closer to the unit circle, leading to slower (longer) transient responses.
\subsection{Pull-Out Probability}\label{ssec:popAnalysis}
    For a GPS receiver, narrowing the tracking loop bandwidth reduces the probability of losing lock due to thermal noise, but it concurrently increases tracking errors caused by vehicle dynamics. This trade-off can be quantified by the pull-out probability (POP). Following~\cite{inproceedings:par2008IONpop}, we develop the POP analysis in three steps: a loss-of-lock criterion, a per-epoch failure probability, and a joint transient-window correction that accounts for the mKF's prolonged transient. We then make explicit how the hardware correlator delay considered in this work enters the analysis, and apply the resulting expression to justify the selection of the correlation time $T = 5$\,ms used throughout this work.

    \subsubsection*{Loss-of-lock criterion}

    A pull-out event is difficult to define rigorously; however, defining it based on the condition most likely to cause it serves as a useful approximation. For the UFA-assisted loop, the boundary of the phase discriminator's linear range provides a suitable criterion. We therefore define a loss-of-lock event to occur when the phase error difference between consecutive epochs exceeds the discriminator's pull-in range~\cite{inproceedings:par2008IONpop}.
    \begin{equation}\label{eq:linearRangeUFA}
        | \Delta\phi_{k} - \Delta\phi_{k-1} | > \frac{\pi}{2} 
    \end{equation}

    \subsubsection*{Per-epoch failure probability}

    Assuming the loop was in lock up to instant $k-1$, the probability that the loss-of-lock condition occurs at instant $k$ is:
    \begin{equation}\label{eq:Pop1}
        P_k = P\left\{ |\Delta\phi_k + n_{\phi(k)} - \Delta\phi_{k-1} - n_{\phi(k-1)}| > \pi/2\right\} \ .
    \end{equation}
    Define the deterministic per-epoch phase-error difference $\mu_k = a\,T^2\,\Delta h(k)$, with $\Delta h(k)$ the phase-error response of the loop to a unit acceleration step at epoch $k$. At the peak epoch, $\mu_{k^*} = K_u a T^2$. Following \cite{inproceedings:par2008IONpop}, the per-epoch failure probability at operating point $\mu_k$ can be approximated as,
    \begin{equation}\label{eq:PopFinalQfuncPAR}
        P_k(\mu_k) \approx Q\left(\sqrt{\rho_k}\right) , \qquad
        \rho_k = \frac{TC/N_0\cos^2(\mu_k)}{1 + TB_N^{'}\sin^2(\mu_k)} \ ,
    \end{equation}
    where $\rho_k$ is the effective per-epoch SNR and $B'_N$ is the equivalent noise bandwidth of the loop cascaded with a differentiator. For the proposed mKF, both $\mu_k$ and $B'_N$ are computed from the closed-loop transfer function $T(z)$ of \eqref{eq:phizAndTz} (that accounts for the delay introduced in previous sections), using \eqref{eq:kfBnT} for $B'_N$.
    
    \subsubsection*{Joint transient-window probability}

    Eq.~\eqref{eq:PopFinalQfuncPAR} was derived in~\cite{inproceedings:par2008IONpop} for a DPLL, where the acceleration-step response is short enough that the peak phase error occurs at essentially a single epoch. The mKF exhibits longer transients due to the inclusion of clock dynamics in the model (Fig.~\ref{fig:rootLocusmKF}), and the loop remains in a near-peak state for multiple consecutive epochs. We account for this by evaluating the joint failure probability over a window of $W$ consecutive epochs centered on the peak epoch~$k^*$.

    Define $\delta_k = \mu_k + \eta_k$, where $\eta_k = n_{\varphi}(k) - n_{\varphi}(k-1)$ is the differenced discriminator noise, and the survival event $S_k = \{|\delta_k| \leq \pi/2\}$. The effective discriminator noise variance at epoch~$k$ is obtained from \eqref{eq:PopFinalQfuncPAR}:
    \begin{equation}\label{eq:sig2k}
        \sigma_k^2 \approx \frac{1}{2\,\rho_k} \ ,
    \end{equation}
    which equals the base atan2 discriminator variance $1/(2T\,C/N_0)$ at $\mu_k = 0$ and increases with $|\mu_k|$ through the squaring-loss factor. Treating the discriminator noise $n_{\varphi}(k)$ as epoch-independent with variance~$\sigma_k^2$, the differenced noise has a tridiagonal covariance:
    \begin{equation}\label{eq:covEta}
        \text{Cov}(\eta_i, \eta_j) = \begin{cases}
            \sigma_i^2 + \sigma_{i-1}^2 & i = j \\
            -\sigma_{\min(i,j)}^2 & |i-j| = 1 \\
            0 & \text{otherwise}
        \end{cases}
    \end{equation}

    Let $\mathcal{W} = \{k^* - \lfloor W/2 \rfloor, \ldots, k^* + \lceil W/2 \rceil - 1\}$ denote the window. The POP is:
    \begin{align}\label{eq:PopJoint}
        P^{\text{KF}} &= 1 - \Pr\!\left(\bigcap_{k\in\mathcal{W}} S_k\right) \\ &= 1 - \int_{-\pi/2}^{\pi/2} \!\cdots\! \int_{-\pi/2}^{\pi/2} \mathcal{N}\!\left(\boldsymbol{\delta};\, \boldsymbol{\mu}_{\mathcal{W}},\, \boldsymbol{\Sigma}_{\mathcal{W}}\right) d\boldsymbol{\delta} \nonumber
    \end{align}
    where $\boldsymbol{\mu}_{\mathcal{W}}$ and $\boldsymbol{\Sigma}_{\mathcal{W}}$ are the mean vector and covariance matrix restricted to $\mathcal{W}$, and $\boldsymbol{\Sigma}_{\mathcal{W}}$ is constructed from \eqref{eq:covEta}. By construction, $P^{\text{KF}}(W)$ is monotonically increasing in $W$. Increasing $W$ captures more of the transient risk but also accumulates the error introduced by the epoch-independence assumption underlying~\eqref{eq:covEta}; the window width $W = 5$ is selected as the value that best balances these two effects, matching the Monte Carlo POP across both the 5\,g and 20\,g scenarios. For a DPLL with no clock-dynamics modeling, the transient collapses to a single epoch, recovering \eqref{eq:PopFinalQfuncPAR} as a special case.

    \subsubsection*{Selection of correlation time $T$}
\begin{figure}
        \centering
        \includegraphics[width=0.49\textwidth]{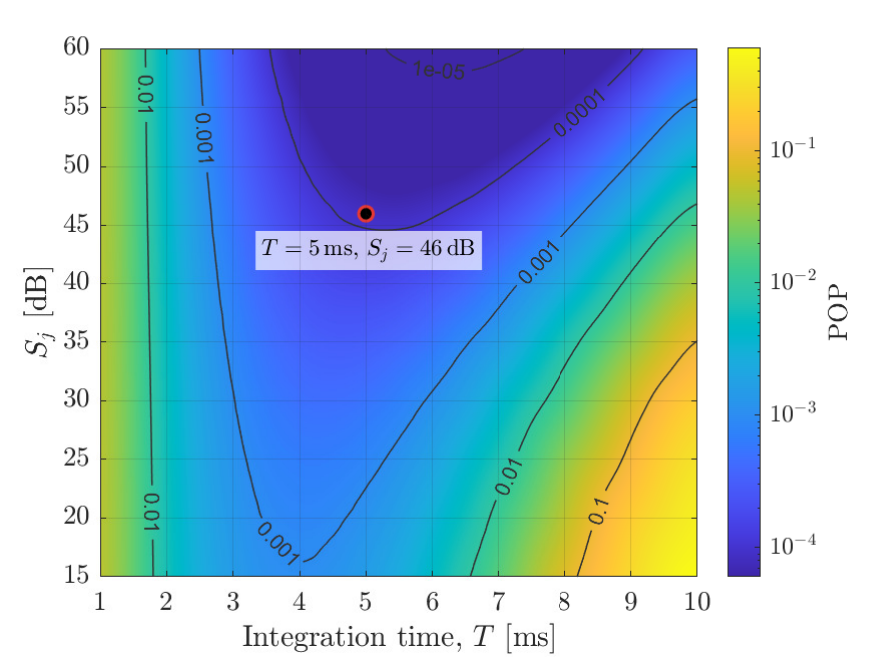}
        \caption{Theoretical POP ($P^{\text{KF}}$) as a function of integration time $T$ and process-noise tuning $S_j$, at $C/N_0 = 34$\,dB-Hz with a 20\,g acceleration step. The marker indicates the operating point at $(T = 5\,\text{ms}, S_j = 46\,\text{dB})$, matching the result of Fig.~\ref{fig:POP_20gACC} and Fig.~\ref{fig:popCompareSIMvsTEO}.}
        \label{fig:popVsTandSj}
    \end{figure}

    The analytical POP across the full $(T, S_j)$ design space at $C/N_0 = 34$\,dB-Hz and a 20\,g acceleration step (see Fig.~\ref{fig:popVsTandSj}) reaches its minimum in the $T = 4$--$8$\,ms range. For a given $T$, a smaller $S_j$ yields a narrower bandwidth and better thermal-noise rejection; the low-POP valley in Fig.~\ref{fig:popVsTandSj} shows that $T = 5$\,ms offers the best trade-off between dynamic robustness and noise performance across the full $S_j$ axis. This value also fulfills the constraint of $T$ being a divisor of the 20\,ms C/A code period, and is consistent with established practice in high-dynamics GNSS tracking~\cite{article:par2012}.

   \subsubsection*{Role of the hardware correlator delay}

    The mKF formulation of Section~\ref{sec:KFbasedTrk}-\ref{ssec:mKF} yields the modified output matrix \eqref{eq:H1r} and closed-loop transfer function \eqref{eq:phizAndTz}, which differ from those of a DPLL or a traditional direct-state KF. The delay modifies both the transient shape $\mu_k$ and the differentiator bandwidth $B'_N$. In a DPLL, the transient peak is concentrated at essentially one epoch, and the single-epoch formula of~\cite{inproceedings:par2008IONpop} suffices. The mKF's pole location (Fig.~\ref{fig:rootLocusmKF}) produces a longer transient in which the phase-error difference remains near its peak for several consecutive epochs, making the single-epoch analysis insufficient. This is the motivation for the joint probability formulation~\eqref{eq:PopJoint}: the per-epoch noise model is inherited from~\cite{inproceedings:par2008IONpop} through $\sigma_k^2 = 1/(2\rho_k)$, but the multi-epoch windowed evaluation is a direct consequence of the hardware delay and its effect on the transient geometry.

    \subsubsection*{Validation through Monte Carlo simulation}

\begin{figure}
        \subfloat[Contour levels for an acceleration step of $5$\,g.]{
            \includegraphics[width=0.49\textwidth]{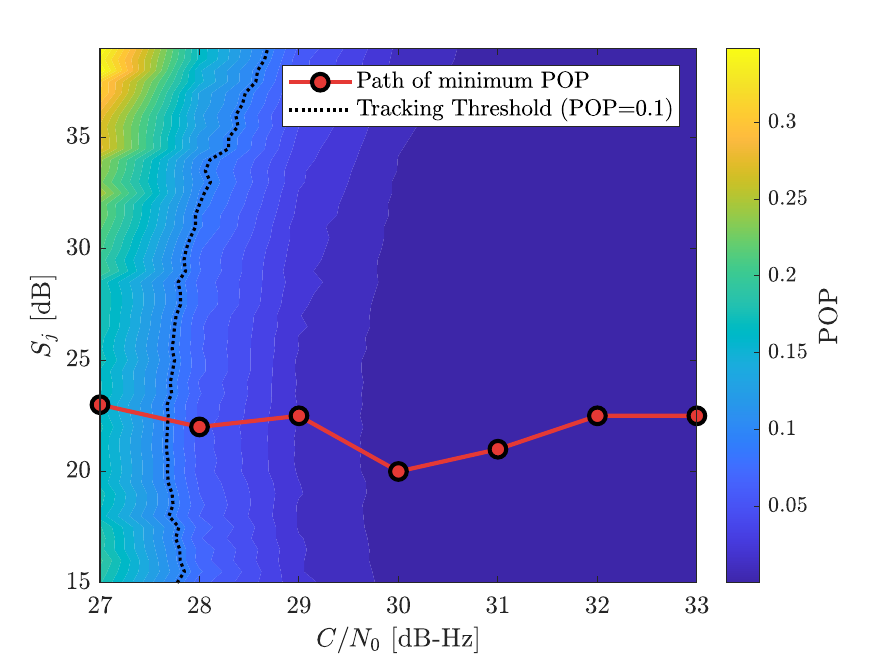}
            \label{fig:POP_5gACC}
        }

        \subfloat[Contour levels for an acceleration step of $20$\,g.]{
            \includegraphics[width=0.49\textwidth]{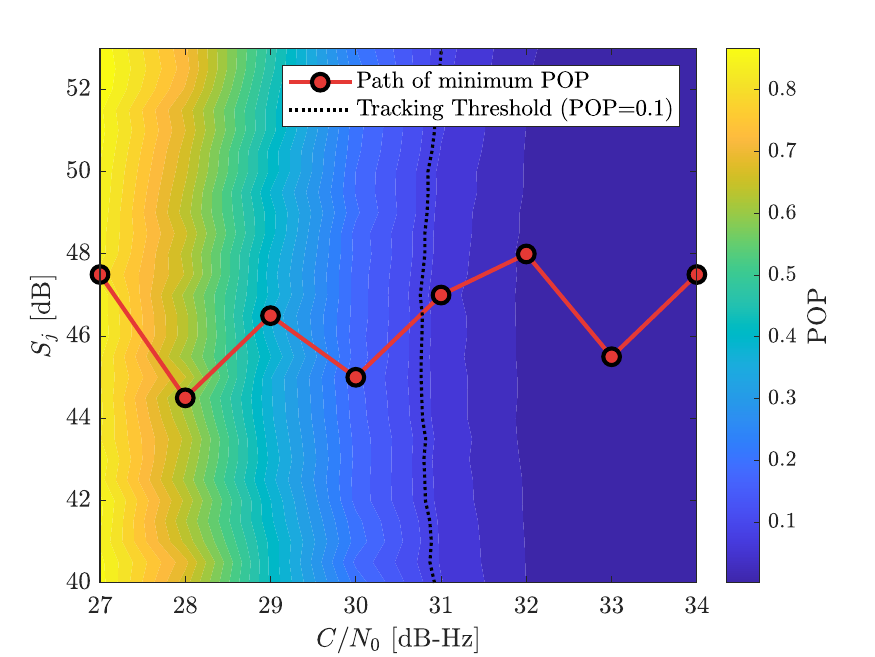}
            \label{fig:POP_20gACC}
        }
        \caption{Simulated POP of the mKF, tracking threshold and minimum probability path.}
        \label{fig:POPcontour}
    \end{figure}

\begin{figure}
        \centering
        \includegraphics[width=0.49\textwidth]{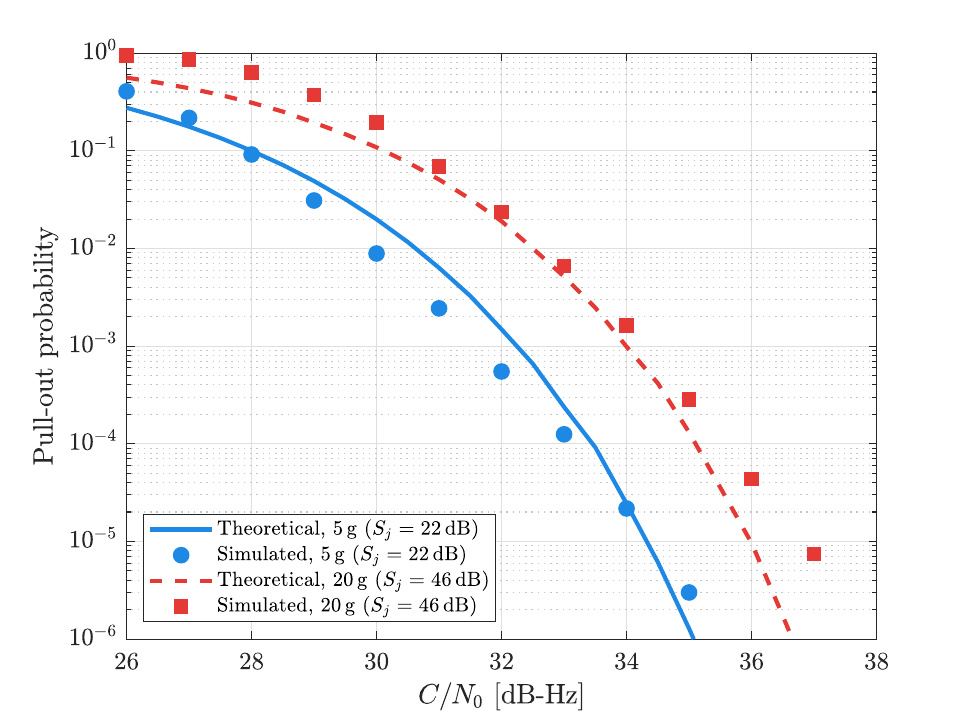}
        \caption{Comparison of the theoretical and simulated POP ($W = 5$ epochs centered on the transient peak).}
        \label{fig:popCompareSIMvsTEO}
    \end{figure}

    The analytical POP expression is validated through Monte Carlo simulation. For each $(C/N_0, S_j)$ pair, $10^7$ independent one-second runs were performed. Each run proceeds as follows: the mKF is initialized in steady state (using the converged Kalman gains for the given $S_j$ and $C/N_0$) with initial state estimates drawn from a Gaussian distribution $\mathcal{N}(\mathbf{0}, \mathbf{P}_{\infty}^+)$, where $\mathbf{P}_{\infty}^+$ is the steady-state \mbox{a-posteriori} error covariance matrix, and correlation values are generated from \eqref{eq:corr1mKF} with thermal noise included. An acceleration step of the specified magnitude is applied at $t = 20$\,ms through phase and frequency profiles that also account for clock dynamics based on \eqref{eq:QmatrixAVAR}. A loss-of-lock event is registered if the condition of \eqref{eq:linearRangeUFA} is met at any epoch during the run, and the POP is estimated as the fraction of runs in which loss-of-lock occurred. Figure~\ref{fig:POPcontour} shows the simulation results, exhibiting a clear valley of optimal performance with an optimal $S_j$ for each $C/N_0$. The tracking threshold, defined as the $C/N_0$ yielding a POP of $0.1$ in one second of tracking, is $28$\,dB-Hz for a 5\,g step and $30.7$\,dB-Hz for a 20\,g step. Figure~\ref{fig:popCompareSIMvsTEO} validates the analytical expression in the operating regime above the threshold within an acceptable match margin in the whole $C/N_0$ range for receiver design purposes.

\subsection{Validation with GNSS Signal Simulator}\label{ssec:validationGSG8samples}
    Selecting the $S_j$ that minimizes the POP for the expected dynamic stress is our definition of optimal tuning. As recorded IF data from real launch campaigns were not accessible to us, the mKF algorithm was validated in a realistic signal environment using a Safran \mbox{GSG-8} GNSS signal generator running the Skydel simulation engine~\cite{misc:safranGSGmanual} to generate GPS L1 C/A signals. The scenario comprises $1$\,s of static operation, a $10$\,s acceleration step (infinite-jerk model, although realized with $20$\,ms settling time) representing engine ignition and cut-off, and $4$\,s of constant-velocity flight. 

    In this validation, the baseband output samples produced by the simulator are post-processed directly by the mKF, without passing through the SENyT receiver's RF front-end used in Section~\ref{sec:mKF}-\ref{ssec:hardwareTesting}. Consequently, the oscillator that drives the sampling and signal generation stage is the simulator's internal reference rather than the TCXO of Section~\ref{sec:signalModel}. The GSG-8 reference is representative of a high-quality oven-controlled crystal oscillator (OCXO), characterized by AVAR parameters $h_0 = 4.5\times10^{-24}$ and $h_{-2} = 4.3\times10^{-23}$~\cite{misc:safranGSGmanual}. To preserve the optimality of the tuning, the POP analysis of Section~\ref{sec:mKF}-\ref{ssec:popAnalysis} was re-evaluated using these OCXO parameters, and the resulting optimal value $S_j = 49$~dB for a $20$\,g acceleration step is the one used in this subsection.

\begin{figure}
        \centering
        \includegraphics[width=0.49\textwidth]{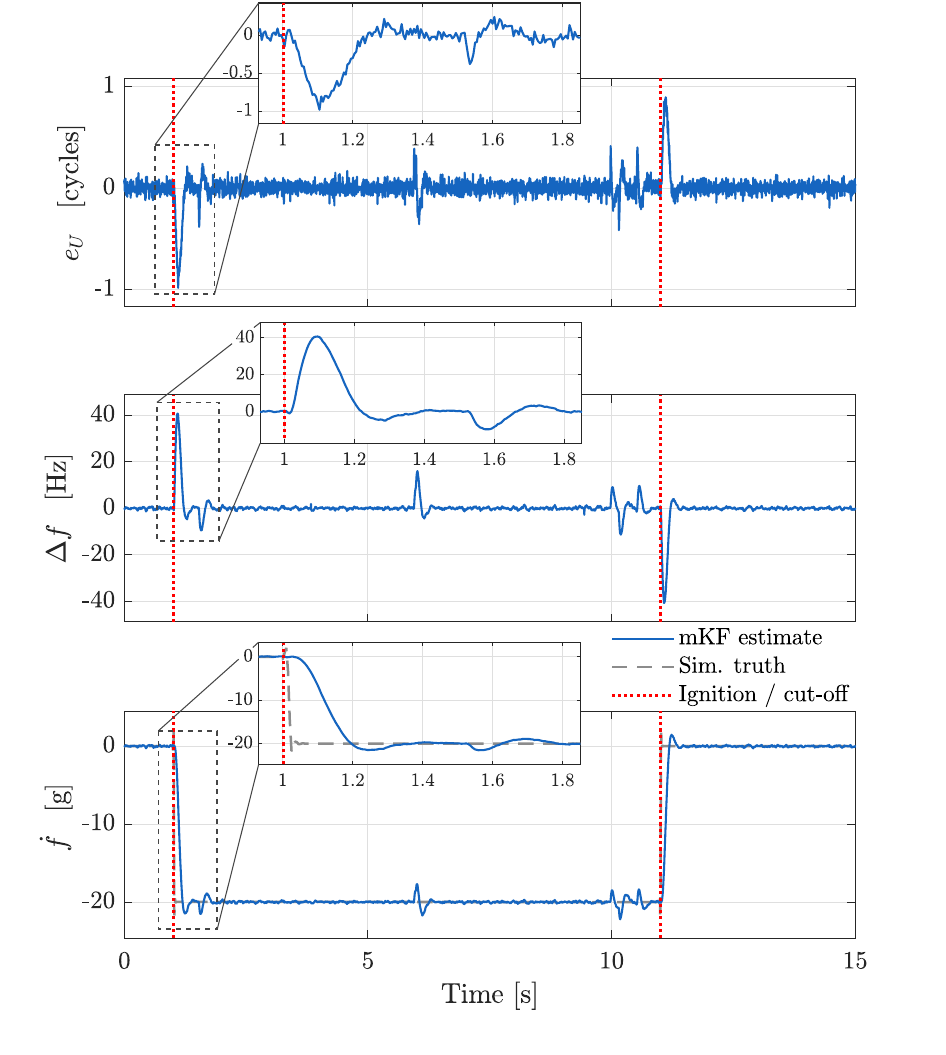}
        \caption{GPS L1 C/A base-band samples phase estimation error and frequency states estimates with mKF. $C/N_0 = 32$\,dB-Hz, $S_j = 49$\,dB, $20$\,g acceleration step.}
        \label{fig:simulatorIFsignalTracking_20g}
    \end{figure}

\begin{figure}
        \centering
        \includegraphics[width=0.49\textwidth]{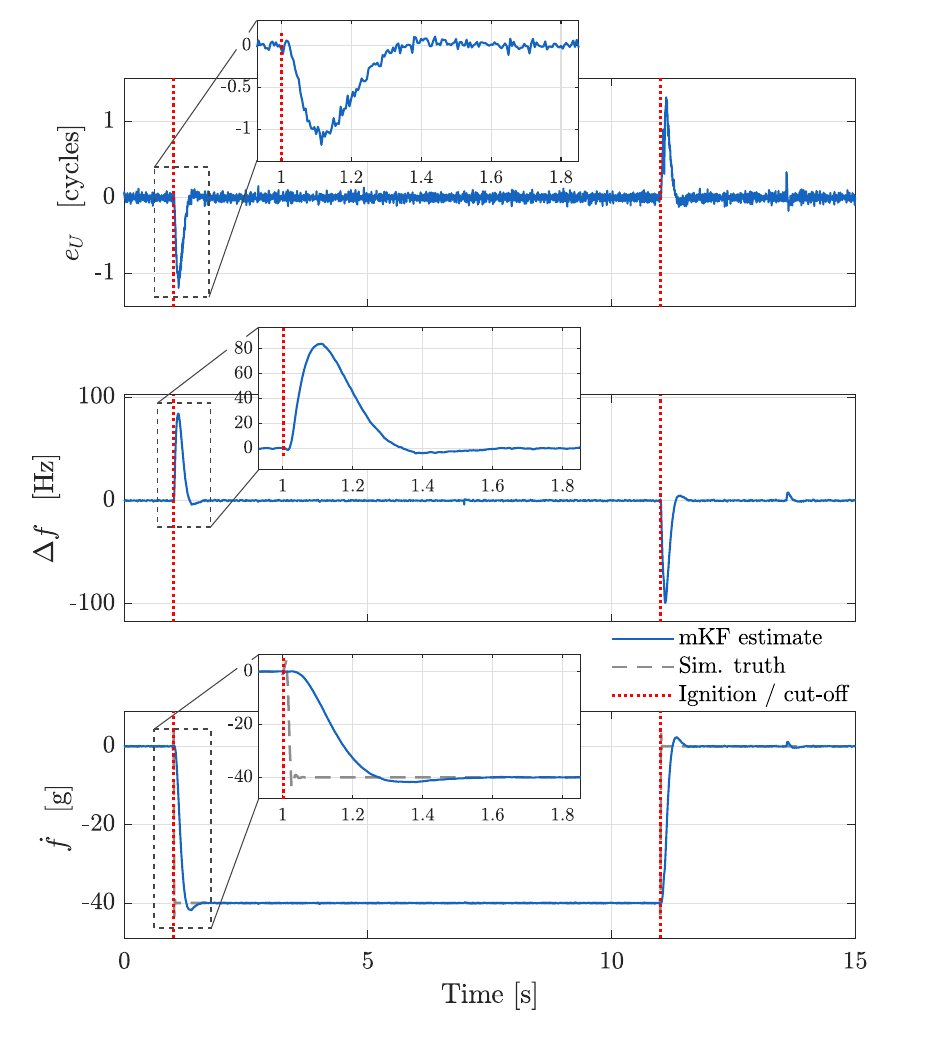}
        \caption{GPS L1 C/A base-band samples phase estimation error and frequency states estimates with mKF. $C/N_0 = 34$\,dB-Hz, $S_j = 49$\,dB, $40$\,g acceleration step.}
        \label{fig:simulatorIFsignalTracking_40g}
    \end{figure}

    The simulator's baseband output samples are post-processed by the steady-state mKF algorithm at $T=5$\,ms with UFA phase discrimination. The frequency state is initialized from the acquisition-stage hint, with all remaining states set to zero. Code tracking is maintained throughout by a standard first-order carrier-aided DLL~\cite{book:kaplan2006}. The resulting phase and frequency estimates for a $20$\,g acceleration step at $C/N_0 = 32$\,dB-Hz are shown in Fig.~\ref{fig:simulatorIFsignalTracking_20g}. This operating point lies only $1.3$\,dB above the tracking threshold for this scenario (see Fig.~\ref{fig:POP_20gACC}), placing the filter under simultaneous dynamic and near-threshold signal stress. Nevertheless, lock is maintained through both acceleration events (ignition and cut-off). The transient overshoots observed during the acceleration step are consistent with discriminator noise amplification near the tracking threshold, and are not indicative of cycle slip; the filter recovers correctly in both cases. A complementary scenario at $40$\,g and $C/N_0 = 34$\,dB-Hz is presented in Fig.~\ref{fig:simulatorIFsignalTracking_40g}. Here the filter is deliberately tuned for $20$\,g dynamics, half the actual acceleration, to assess robustness against underdimensioned tuning. Despite this mismatch, the filter maintains lock throughout, demonstrating that the proposed algorithm tolerates significant dynamic underestimation without loss of lock. Notably, the phase error $e_U$ briefly exceeds one cycle without producing a cycle slip; this is a direct consequence of the UFA discriminator's extended linear range beyond that of the standard arctangent detector (see condition in~\eqref{eq:linearRangeUFA}). Tuning the filter with the $S_j$ that minimizes the POP for the true $40$\,g dynamic would reduce this transient. Together, both figures confirm that the proposed tracking algorithm successfully handles acceleration steps of magnitudes up to at least 40\,g.

\subsection{Real-time Hardware Validation}\label{ssec:hardwareTesting}
    As a final validation step, the proposed mKF was implemented as part of the tracking software of a real-time GPS receiver with hardware correlators developed by SENyT (UNLP)~\cite{inproceedings:SR2017URUCONqseries,article:qseries2020Sr}. The algorithm was coded in C and executed on a LEON3 processor running the RTEMS real-time operating system, embedded in a Field-Programmable Gate Array (FPGA) that also implements the acquisition and tracking correlation channels for the GPS L1 C/A signal. The tracking software takes in the correlator outputs at each epoch and configures the hardware replica generators with the resulting state estimates. Under this architecture, the processing and correlator delays modeled in software in the previous sections become intrinsic to the implementation, providing the natural setting for which the mKF was designed.

    The oscillator that drives the receiver's RF front-end is the same TCXO whose AVAR parameters were adopted throughout the design and characterization of the mKF (see Section~\ref{sec:signalModel}). Therefore, the optimal $S_j$ values derived from the POP analysis of Section~\ref{sec:mKF}-\ref{ssec:popAnalysis} for the design TCXO are directly applicable to this experiment without re-tuning: $S_j = 22$\,dB for a $5$\,g acceleration step and $S_j = 46$\,dB for a $20$\,g step.

    The validation scenario uses the GSG-8 hardware signal generator with the RF output connected to the receiver's antenna input, running the Skydel simulation engine. A GPS L1 C/A signal, quantized at 2-bit resolution after the front-end stage, is tracked in real time. Once the filter reaches steady state in a static scenario, the high-dynamic profile described in Section~\ref{sec:mKF}-\ref{ssec:validationGSG8samples} is applied.

    To isolate the effect of the proposed delay treatment, a second KF variant was implemented on the same receiver. Both variants are tuned with the same $S_j$ for each scenario and operate on the same RF signal:
    \begin{itemize}
    \item \textbf{mKF} (proposed): uses $\mathbf{H} = \bm{e}_1\mathbf{F}^{-1}$ and rotates the correlation output with $\hat{\phi}_{k-1}^+$, as described in Section~\ref{sec:KFbasedTrk}-\ref{ssec:mKF}.
    \item \textbf{tradKF} (delay-unaware): a direct-state KF with $\mathbf{H} = \bm{e}_1$ that does not account for the processing delay. The correlator is configured with $\hat{\mathbf{x}}_k^-$ and the correlation has a phase reference $\hat{\phi}_{k}^-$, as in a standard KF implementation that assumes zero-delay correlators.
    \end{itemize}

\begin{figure}
        \subfloat[Phase error $e_U$ and frequency error $\Delta f$ at $5$\,g, $S_j = 22$\,dB]{
            \includegraphics[width=0.49\textwidth]{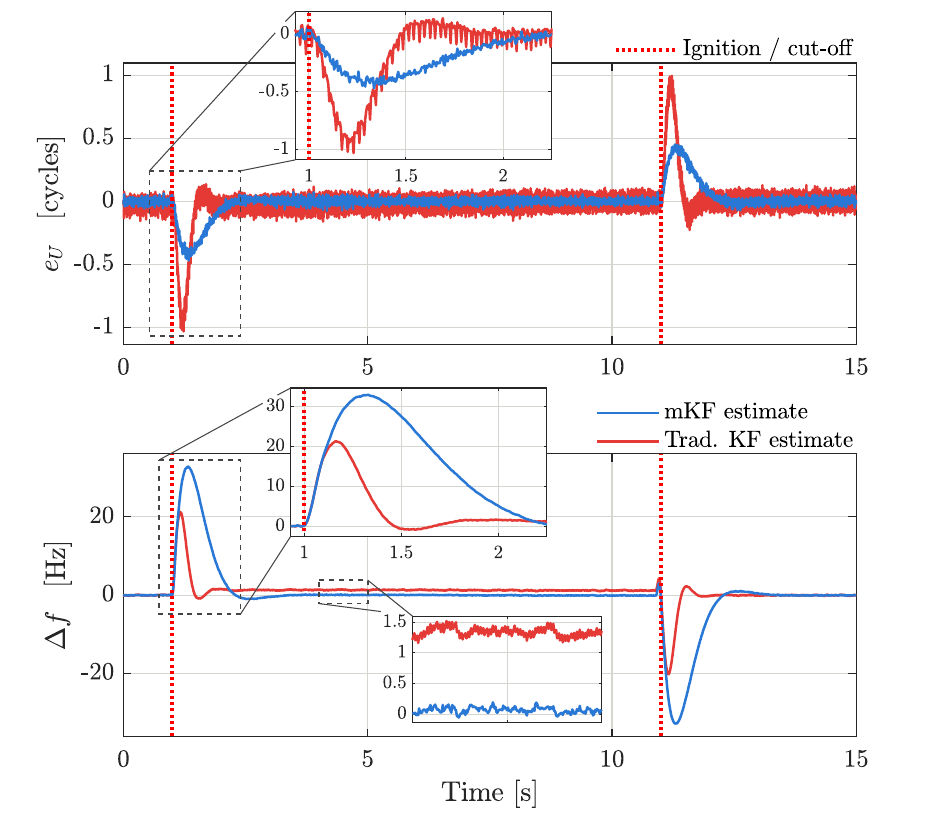}
            \label{fig:HW_AB_5g}
        }

        \subfloat[Phase error $e_U$ and frequency error $\Delta f$ at $20$\,g, $S_j = 46$\,dB]{
            \includegraphics[width=0.49\textwidth]{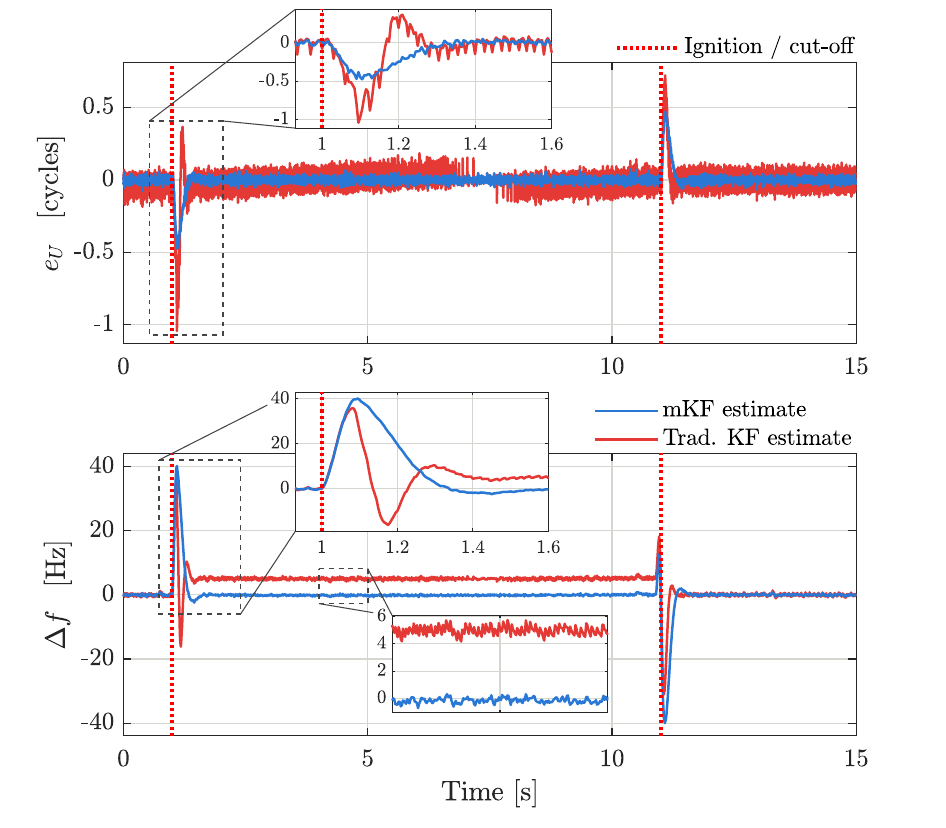}
            \label{fig:HW_AB_20g}
        }
        \caption{Controlled comparison on the SENyT receiver at $C/N_0 = 37$\,dB-Hz: proposed mKF vs.\ delay-unaware traditional KF.}
        \label{fig:HW_AB_comparison}
    \end{figure}

    The resulting phase and frequency error sequences are shown in Figs.~\ref{fig:HW_AB_5g} and~\ref{fig:HW_AB_20g} and compared to the delay-unaware tradKF. In both scenarios the mKF maintains lock throughout the complete engine ignition and cut-off transient, with phase-error excursions bounded within the extended UFA linear range and returning to the steady-state regime after each event. The tradKF, in contrast, exhibits two distinct pathologies.
 
    First, the tradKF produces significantly larger phase-error transients: approximately $0.9$\,cycles at $5$\,g (compared to ${\sim}0.4$\,cycles for the mKF) and approaching $1$\,cycle at $20$\,g, reducing the margin to the UFA discriminator's pull-in boundary and making the filter susceptible to cycle slips under repeated runs. Second, the zoomed frequency-error panels reveal a persistent non-zero steady-state frequency offset, approximately $1$--$1.5$\,Hz at $5$\,g and $5$--$6$\,Hz at $20$\,g, while the mKF settles to zero. This bias is the direct consequence of ignoring the one-epoch temporal misalignment between the local replica and the incoming signal: the filter's measurement model assumes an alignment that is different from the one existing in the hardware, and the frequency state absorbs the resulting systematic innovation error as a constant offset that scales with the acceleration magnitude.
 
    Beyond the transient peaks and the frequency bias, the tradKF's phase-error trace exhibits a structured, low-frequency variation that is clearly distinct from thermal noise, particularly visible in the $20$\,g scenario between the ignition and cut-off events. This behavior is driven by the same temporal misalignment: the uncorrected frequency offset continuously feeds back through the loop, modulating the phase error with a frequency-dependent component that the filter cannot attribute to any modeled noise source. The mKF's phase error, in contrast, presents the expected thermal-noise-dominated structure throughout the steady-state regime.

    The larger transient peaks observed for the tradKF reduce the margin to the discriminator's pull-in boundary, making the filter more susceptible to cycle slips under repeated runs. The persistent frequency bias, while not causing immediate loss of lock in these realizations, degrades the quality of the frequency and code-tracking aiding, and would accumulate additional positioning error in a navigation solution. These results demonstrate that accounting for the processing delay is necessary for reliable carrier tracking on a receiver with hardware correlators.

\begin{figure}
        \centering
        \includegraphics[width=0.49\textwidth]{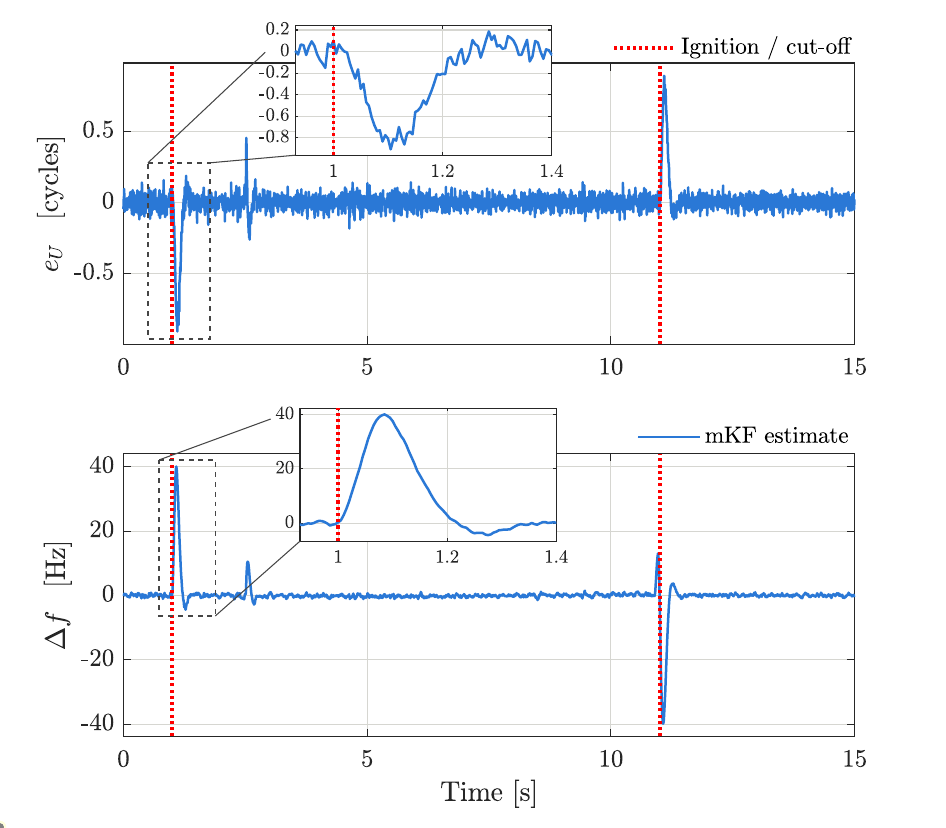}
        \caption{Real-time GPS L1 C/A tracking with the proposed mKF at $C/N_0 = 32$\,dB-Hz, $S_j = 46$\,dB, $20$\,g acceleration step.}
        \label{fig:HW_nearTTh}
    \end{figure}

    To further demonstrate that the mKF operates reliably near the tracking threshold on hardware, Fig.~\ref{fig:HW_nearTTh} shows the mKF tracking at $C/N_0 = 32$\,dB-Hz with a $20$\,g acceleration step and $S_j = 46$\,dB. This operating point lies approximately $1.3$\,dB above the analytical tracking threshold for this scenario, placing the filter under simultaneous dynamic and near-threshold signal stress. The transient peak reaches approximately $0.8$\,cycles, consistent with the increased discriminator noise at low $C/N_0$. Nevertheless, the mKF maintains continuous lock throughout both the acceleration onset and cut-off transients, and the frequency error recovers cleanly to zero.

    These results confirm the practical effectiveness of the proposed mKF under real-time operating conditions, with all processing and correlator delays intrinsically present, and validate that the minimum POP tuning methodology yields the expected tracking behavior when deployed on the target hardware.
\section{COMPARISON BETWEEN MODIFIED KALMAN FILTER AND ONE-DELAY DPLL}\label{sec:KFvsPLL}
    The design of optimal loop filters for DPLLs is a well-addressed field. Given that it is the standard architecture for carrier tracking loops due to its simplicity, various methods have been proposed. An optimal design based on analog principles was presented in \cite{article:jaffeRechtin1955nearoptimumDLLs}, which solves the well-known bandwidth \mbox{trade-off} by minimizing a functional that weighs output noise against the transient response energy. To overcome the bandwidth restrictions of that method, \cite{article:stephenThomas1995ControlledRootDPLL} introduced a digital design based on pole placement, though it still relied on analog design rules to determine the pole locations. A fully digital model was later formalized in \cite{inproceedings:par2007ION}, which minimized the quadratic functional directly in the digital domain, achieving loops capable of tracking dynamics of up to $40$\,g.

    This section establishes the equivalence between the proposed mKF and these optimal DPLL designs. The relationship between steady-state KFs and constant-bandwidth DPLLs has been established in \cite{article:driessen1994} and more generally, for loops of any order, in \cite{article:bidonRoche2024equivalenceKFandDPLL}. On the other hand, an explicit equivalence between DPLL filter taps and KF gains is presented in \cite{article:1yadeMorton2017generalizedGnssSignalTRK,article:shu2013ThirdOrderKalmanFilter}, while \cite{article:2yadeMorton2017generalizedGnssSignalTRK} analyzes its performance. However, those approaches still rely on discretizing an analog PLL model. Thus, we will develop this equivalence entirely within the digital domain.
\subsection{From mKF Gains to DPLL Filter Taps}
    The work in \cite{article:par2012} presents an optimal loop filter for tracking an acceleration step. This filter is derived from a one-sample optimal smoothing problem, which is mathematically equivalent to the one-delay DPLL filter of interest here. The loop filter is given by
    \begin{equation}\label{eq:dpllLoopFilter}
        F(z) = \frac{(3-p_s)+(p_d-3)z^{-1} + (1-p_p)z^{-2}}{(1-z^{-1})^3} ,
    \end{equation}
    where $p_s = p_0 + p_1 + p_2$, $p_d = p_0p_1 + p_0p_2 + p_1p_2$ and $p_p = p_0p_1p_2$. These poles ($p_i$) are determined by the single design parameter $\nu$, which selects the loop bandwidth. The value of $\nu$ depends on the acceleration step magnitude, the correlation time, and a weighting factor that explicitly controls the trade-off between minimizing output noise variance and minimizing the energy of the transient response. As presented in \cite{article:par2012} the poles can be obtained through \mbox{$p_{1,2} + p_{1,2}^{-1} = 2-\frac{1 \pm j\sqrt{3}}{2}\sqrt[3]{\nu}$} and $p_{3} + p_3^{-1} = 2+\sqrt[3]{\nu}$.

    By comparing the open-loop transfer function of the mKF from~\eqref{eq:Nz} with this optimal DPLL filter, we find a direct structural equivalence within a purely digital framework. However, it is important to note that if the mKF is tuned to account for clock dynamics (using $\mathbf{Q}_A$), its pole locations will differ from those obtained using the optimization in \cite{article:par2012}, as that DPLL optimization does not consider clock noise. Thus, the filter structures are equivalent but their time responses differ with the tuning and noise model.

    The design bandwidth accurately reflects the resulting digital bandwidth, without the limitations of analog-to-digital conversions. However, the validity of the underlying steady-state models relies on the standard linearization assumptions inherent in both KF analysis and the DPLL derivation. This equivalence is expected, as both the steady-state KF and the optimal DPLL in \cite{article:par2012} derive from Wiener filtering for the identical one-delay architecture.
\subsection{Comparative Tuning and Performance Analysis}\label{sec:comparativeTuningAndPerformanceAnalysis}
    As mentioned previously, the Kalman gains cannot be arbitrarily placed to match any desired optimal one-delay DPLL response. As we aim to compare and analyze the performance of both architectures, we propose a tuning criterion for the mKF by setting an equal $B_N$ for the two filters.

\begin{figure}
        \centering
        \includegraphics[width=0.49\textwidth]{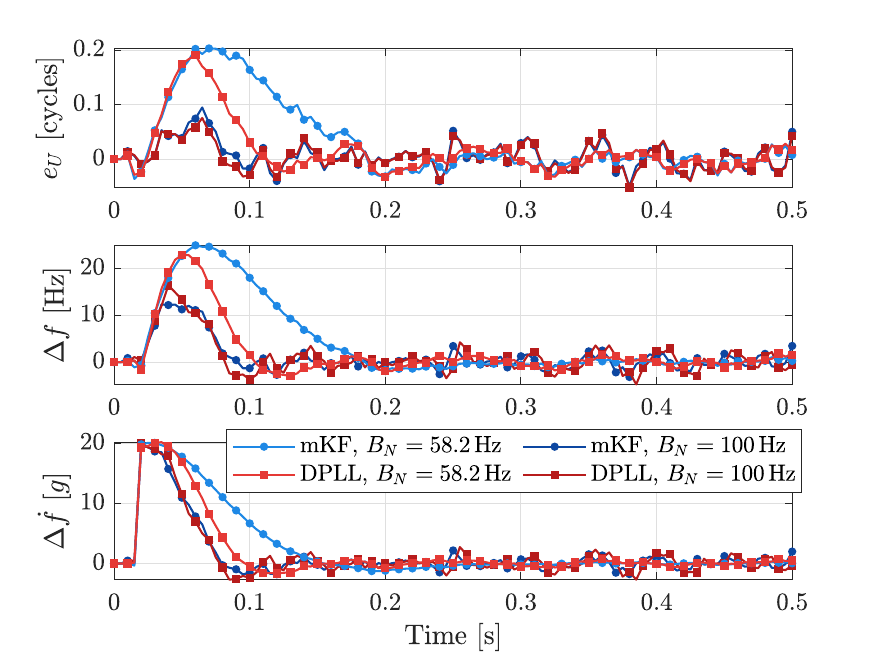}
        \caption{State estimation error for mKF and DPLL. $58.2$\,Hz and $100$\,Hz bandwidths, $20$\,g acceleration step and $C/N_0 = 40$\,dB-Hz.}
        \label{fig:CompareKfDPLL}
    \end{figure}

\begin{figure}
        \centering
        \includegraphics[width=0.49\textwidth]{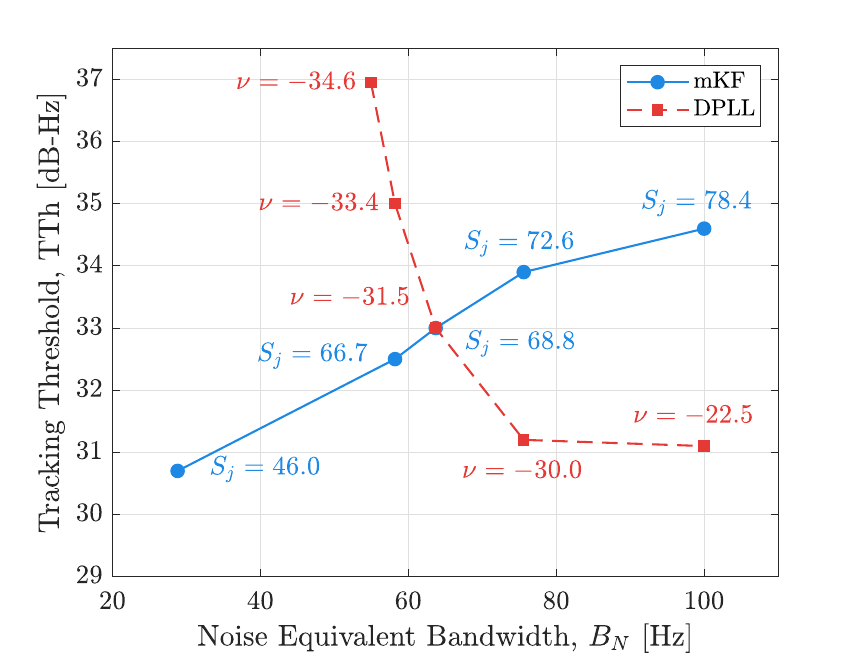}
        \caption{KF and DPLL tracking thresholds for $20$\,g acceleration step. The mKF was tuned assuming a design $C/N_0=34$\,dB-Hz. Datapoint values are displayed in dB.}
        \label{fig:tthKFvsDPLL}
    \end{figure}

    To compare their performance, we consider two representative bandwidths: $58.2$\,Hz, a standard value for high-dynamics tracking, and $100$\,Hz. Both configurations are achieved with the mKF by setting $S_j = 56.4$\,dB and $S_j = 71.5$\,dB, respectively (at $C/N_0 = 40$\,dB-Hz). A comparison at equal bandwidths is fair because it ensures both loops achieve the same phase error variance and exhibit similar dynamic response characteristics. The time response of a single realization at both bandwidths is shown in Fig.~\ref{fig:CompareKfDPLL} (with the DPLL tuned as in Fig.~\ref{fig:tthKFvsDPLL}), and the steady-state standard deviations averaged over 1000 independent runs (discarding the transient) are reported in Table~\ref{tab:EstimationErrors}. While both architectures achieve equal variance for the phase error at each bandwidth, the mKF achieves lower estimation error variance for the higher-order states (frequency and frequency rate) in both cases. This is an expected result, as the KF is by definition the optimal estimator in the MMSE sense for the complete state vector, while the DPLL only optimizes the phase state.
    
    Figure~\ref{fig:CompareKfDPLL} also allows a direct comparison of the transient responses at high $C/N_0$. At $B_N = 58.2$\,Hz, the mKF exhibits a slower transient than the DPLL, with a settling time of approximately $0.15$\,s compared to $0.10$\,s. This behavior is expected, since modeling the clock dynamics shifts the mKF's dominant poles closer to the unit circle, as anticipated by the root locus analysis in Fig.~\ref{fig:rootLocusmKF}. At $B_N = 100$\,Hz, the transient responses of the two architectures converge: the dominant poles of both filters present similar magnitudes at this bandwidth, and the influence of the clock-dynamics model on the mKF's transient becomes negligible, which reconfirms the behavior anticipated in Fig.~\ref{fig:rootLocusmKF}. Finally, the maximum residual $\Delta f \approx 25$\,Hz visible in Fig.~\ref{fig:CompareKfDPLL} corresponds to $\sim 0.22$\,dB of sinc attenuation at $T = 5$\,ms, validating the $\mathrm{sinc}(\cdot) \approx 1$ assumption at high $C/N_0$ of Section~\ref{sec:signalModel}.
    \begin{table}
        \centering
        \caption{Steady-state estimation error standard deviations. $58.2$\,Hz and $100$\,Hz bandwidths, $20$\,g acceleration step, $C/N_0 = 40$\,dB-Hz.}
        \label{tab:EstimationErrors}
        \setlength{\tabcolsep}{5pt}
        \begin{tabular}{|c|c|c|c|c|c|c|}
            \hline
            \multirow{2}{*}{$B_N$}
                & \multicolumn{2}{|c|}{$\sigma_{\Delta\phi}$ [cycles]}
                & \multicolumn{2}{c|}{$\sigma_{\Delta f}$ [Hz]}
                & \multicolumn{2}{c|}{$\sigma_{\Delta\dot{f}}$ [g]} \\
            \cline{2-7}
                & mKF & DPLL & mKF & DPLL & mKF & DPLL \\
            \hline
            $58.2$\,Hz & 0.014 & 0.014 & 0.45 & 0.81 & 0.14 & 0.34 \\
            $100$\,Hz  & 0.019 & 0.019 & 1.19 & 1.66 & 0.7  & 0.98 \\
            \hline
        \end{tabular}
    \end{table}
    To evaluate sensitivity under dynamic stress, Fig.~\ref{fig:tthKFvsDPLL} compares the tracking threshold (TTh) of the proposed mKF and the optimal one-delay DPLL under a $20$\,g acceleration step across various ($B_N$). The plot reveals distinct operational regimes for both architectures. At low bandwidths, the mKF significantly outperforms the DPLL, achieving its absolute minimum TTh of $30.7$\,dB-Hz at its optimal configuration ($B_N = 28.8$\,Hz). In contrast, the DPLL exhibits a strict lower bandwidth boundary; it becomes unable to accommodate the dynamic stress and fails to maintain lock entirely below $B_N \approx 55$\,Hz. As the bandwidth increases, the DPLL's performance recovers rapidly until the two architectures intersect at $B_N = 63.7$\,Hz, where both require a TTh of $33$\,dB-Hz. Beyond this crossing point, the behaviors diverge. The mKF's required threshold increases steadily alongside the bandwidth due to the wider noise admission. Conversely, the DPLL's threshold flattens into an asymptotic lower limit, eventually reaching a minimum TTh of $31$\,dB-Hz at $B_N = 100$\,Hz. This demonstrates that while the DPLL can eventually approach the peak sensitivity of the mKF, it requires a severely widened bandwidth to do so.

    The tuning of the mKF for Fig.~\ref{fig:tthKFvsDPLL} was done considering a design $C/N_0=34$\,dB-Hz (different from the $C/N_0=40$\,dB-Hz considered in Fig.~\ref{fig:CompareKfDPLL}). It was shown previously that the resulting bandwidth is dependent on $C/N_0$ and to account for this, the threshold results are presented at a nominal operative $C/N_0 = 34$\,dB-Hz. At higher $C/N_0$ the filter operates with increased margin from the tracking threshold, while at lower $C/N_0$ the optimal bandwidth narrows, reducing the effective TTh and thereby preserving the tracking margin claimed in this work.

    Figures~\ref{fig:CompareKfDPLL} and~\ref{fig:tthKFvsDPLL} reveal a complementary advantage of the proposed architecture. At any given bandwidth, the mKF matches the DPLL's phase error variance while achieving lower estimation error in the higher-order states, as confirmed by Table~\ref{tab:EstimationErrors}. At the same time, Fig.~\ref{fig:tthKFvsDPLL} shows that the mKF attains its minimum tracking threshold at $B_N = 28.8$\,Hz, a bandwidth at which the DPLL cannot sustain lock under the same dynamic stress. The DPLL only approaches a comparable sensitivity at $B_N = 100$\,Hz, where it admits substantially more noise, degrading precisely the higher-order state estimates where the mKF holds its structural advantage. This dual benefit, superior state estimation at equal bandwidth, and viable low-bandwidth operation inaccessible to the DPLL, constitutes the practical case for the proposed mKF architecture in high-dynamics aerospace applications.
\section{CONCLUSIONS}\label{sec:conclusions}
    The timing constraints and inherent delays present in real-time Kalman filter-based tracking loops with hardware correlators have been thoroughly analyzed. A novel Modified KF (mKF) formulation was proposed to address these issues, and its equivalence to the optimal Wiener filter was established. This equivalence allowed for the analytical derivation of the filter's closed-loop transfer function, which effectively accounts for the implementation delay. A direct relationship between the filter's poles and its tuning parameters was established through this, enabling the precise computation of the mKF's noise equivalent bandwidth.

    This approach represents an improvement over previous works, which either neglected the delay or relied on analog-to-digital model conversions. However, a practical method for selecting the filter's tuning parameters was still needed. To address this, we derived an analytical expression for the pull-out probability (POP) of the mKF.

    The POP was formulated as the joint probability of the phase-error difference exceeding the discriminator's linear range during the transient, evaluated over a window of epochs centered on the peak of the deterministic response. An epoch-dependent noise model, derived from the per-epoch SNR of~\cite{inproceedings:par2008IONpop}, captures the squaring-loss amplification during the transient. This analytical POP expression was validated through Monte Carlo simulations, which demonstrated its accuracy for both the 5\,g and 20\,g acceleration scenarios without scenario-dependent tuning. Evaluating the analytical POP over the full $(T, S_j)$ design space further provided a principled justification for the correlation time selected throughout this work. The simulated results revealed a clear optimal tuning for each dynamic stress level, establishing a straightforward method to configure the mKF for minimum tracking threshold in high-dynamic scenarios. Moreover, the methodology and the mKF were tested on GNSS signal generator baseband samples, successfully tracking acceleration steps of magnitudes up to at least 40\,g, and on an embedded real-time receiver, where the processing delays addressed in this work are physically present rather than modeled.

    Finally, the proposed mKF was compared with a one-delay DPLL design. An equivalence between the two architectures was established entirely in the digital domain, allowing a direct comparison of filters with equal order and delays. At equal noise bandwidth, the mKF matches the DPLL's phase error variance while achieving lower estimation error in the higher-order states (frequency and frequency rate). Moreover, the mKF sustains lock at bandwidths inaccessible to the one-delay DPLL under the same dynamic stress, attaining its minimum tracking threshold at a bandwidth where the DPLL fails entirely. This dual benefit positions the mKF as a robust and noise-efficient tracking architecture for high-dynamic aerospace GNSS applications.
\section*{Acknowledgment}
The authors would like to thank \textit{Safran Electronics \& Defense} for providing the Skydel GNSS Simulation Software, as part of the Minerva Academic Partnership Program (formerly Orolia Academic Partnership Program).
    \bibliographystyle{IEEEtran}
    \bibliography{Bibliography/IEEEabrv, Bibliography/Articles, Bibliography/Books, Bibliography/InProceedings, Bibliography/Misc}
\end{document}